\RequirePackage{etex}

\documentclass[prb,superscriptaddress,twocolumn,showpacs]{revtex4-1}
\pdfoutput=1
\usepackage{amsmath, amsthm, amssymb, mathrsfs}
\usepackage[pdfborder={0 0 0}]{hyperref}

\usepackage{dcpic, pictex, setspace, rotating}
\usepackage{array, verbatim, bbm, faktor}
\usepackage{graphicx}
\usepackage{times}
\usepackage{pinlabel}
\usepackage{tikz}

\newcommand{\be}{\begin{equation}}
\newcommand{\ee}{\end{equation}}

\newtheorem{theorem}{Theorem}[section]
\newtheorem{lemma}[theorem]{Lemma}

\newtheorem{corollary}[theorem]{Corollary}

\theoremstyle{definition}
\newtheorem{definition}[theorem]{Definition}

\theoremstyle{remark}

\newcommand{\bone}{{\boldsymbol{1}}}
\newcommand{\btau}{{\boldsymbol{\tau}}}
\newcommand{\brho}{{\boldsymbol{\rho}}}

\newcommand{\eg}{\textit{e.g.}\ }
\newcommand{\ie}{\textit{i.e.}\ }

\newcommand{\C}{{\mathbb C}}

\newcommand{\til}{\widetilde}

\newcommand{\mO}{\mathcal O}
\newcommand{\mL}{\mathcal L}

\def\hline{\bigskip\hrule\bigskip}  % temporary def
\newcommand{\ket}[1]{\left|{#1}\right\rangle}
\newcommand{\bra}[1]{\left\langle{#1}\right|}

\begin{document}
\title{Galois Conjugates of Topological Phases}
%\date{\today}

\author{M. H. Freedman}
\affiliation{Microsoft Research, Station Q, University of California, Santa Barbara, CA 93106, USA}

\author{J. Gukelberger}
\affiliation{Theoretische Physik, ETH Zurich, 8093 Zurich, Switzerland}

\author{M. B. Hastings}
\affiliation{Microsoft Research, Station Q, University of California, Santa Barbara, CA 93106, USA}

\author{S. Trebst}
\affiliation{Microsoft Research, Station Q, University of California, Santa Barbara, CA 93106, USA}

\author{M. Troyer}
\affiliation{Theoretische Physik, ETH Zurich, 8093 Zurich, Switzerland}

\author{Z. Wang}
\affiliation{Microsoft Research, Station Q, University of California, Santa Barbara, CA 93106, USA}

\date{\today}

\begin{abstract}
Galois conjugation relates unitary conformal field theories (CFTs) and topological quantum field theories (TQFTs) to their non-unitary counterparts. Here we investigate Galois conjugates of quantum double models, such as the Levin-Wen model. While these Galois conjugated Hamiltonians are typically non-Hermitian, we find that their ground state wave functions still obey a generalized version of the usual code property (local operators do not act on the ground state manifold) and hence enjoy a generalized topological protection. 
The key question addressed in this paper is whether such non-unitary topological phases can also appear as the ground states of Hermitian Hamiltonians. Specific attempts at constructing Hermitian Hamiltonians with these ground states lead to a loss of the code property and topological protection of the degenerate ground states. 
%Beyond this we rigorously prove that the ground states of the Galois conjugated doubled Fibonacci theory cannot be locally conjugated to the ground states of a gapped Hermitian model in a topological phase, and conjecture that a similar statement holds for any non-unitary TQFT. 
Beyond this we rigorously prove that no local change of basis (\ref{theorema}) can transform the ground states of the Galois conjugated doubled Fibonacci theory into the ground states of a topological model whose Hermitian Hamiltonian satisfies Lieb-Robinson bounds. These include all gapped local or quasi-local Hamiltonians. A similar statement holds for many other non-unitary TQFTs. One consequence  is that the ``Gaffnian" wave function cannot be the ground state of a gapped fractional quantum Hall state.
\end{abstract}

% 05.30.Pr – Fractional statistics systems (anyons, etc.)

\maketitle

\section{Introduction}

%\begin{itemize}
%\item topological phases
%\item code property
%\item Galois conjugation
%\item non-unitary TQFTs
%\item Gaffnian FQH wave function
%\item can there be non-unitary topological phases?
%\end{itemize}

{\em Galois conjugation}, by definition, replaces a root of a polynomial by another one with identical algebraic properties.  For example, $i$ and $-i$ 
are Galois conjugate (consider $z^2+1=0$) as are $\phi=\frac{1+\sqrt{5}}{2}$ and $-\frac{1}{\phi}=\frac{1-\sqrt{5}}{2}$ (consider $z^2-z-1=0$), as well as $\sqrt[3]{2}$, $\sqrt[3]{2}e^{2\pi i/3}$, and $\sqrt[3]{2}e^{-2\pi i/3}$ (consider $z^3-2=0$).  In physics Galois conjugation can be used to convert non-unitary conformal field theories (CFTs) to unitary ones, and vice versa.
One famous example is the non-unitary Yang-Lee CFT, which is Galois conjugate to the Fibonacci CFT $(G_2)_1$,
the even (or integer-spin) subset of su(2)$_3$.
%variants of unitary , for example the unitary Fibonacci CFT ``Fib" is Galois conjugated to the non-unitary Yang-Lee CFT ``YL", {\em i.e.} Fib$\rightsquigarrow${YL} 
%(Fib is realized, for example, as $(G_2)_1$).  

In statistical mechanics non-unitary conformal field theories have a venerable history.\cite{YangLee,Cardy}
However, it has remained less clear if there exist physical situations in which non-unitary models can provide a useful description of the low energy physics of a quantum mechanical system  -- after all, Galois conjugation typically destroys the Hermitian property of the Hamiltonian.  
Some non-Hermitian Hamiltonians, which surprisingly have totally real spectrum, have been found to arise in the study of $PT$-invariant one-particle systems \cite{Bender} and in some Galois conjugate many-body systems \cite{YLC} and might be seen to open the door a crack to the physical use of such models.  
Another situation, which has recently attracted some interest, is the question whether non-unitary models can describe 1D edge states of certain 2D bulk states (the edge holographic for the bulk). In particular, there is currently a discussion on whether or not the ``Gaffnian" wave function could be the ground state for a {\em gapped} fractional quantum Hall (FQH) state albeit with a non-unitary ``Yang-Lee" CFT describing its edge.\cite{r09b,r09a,Gaffnian}  We conclude that this is not possible, further restricting the possible scope of non-unitary models in quantum mechanics.

We reach this conclusion quite indirectly.  Our main thrust is the investigation of Galois conjugation in the simplest non-Abelian Levin-Wen model.\cite{LW} This model, which is also called ``DFib", is a topological quantum field theory (TQFT) whose states are string-nets on a surface labeled by either a trivial or ``Fibonacci" anyon.  From this starting point, we give a rigorous argument that the ``Gaffnian" ground state cannot be locally conjugated to the ground state of any topological phase, within a Hermitian model satisfying Lieb-Robinson (LR) bounds~\cite{lr} (which includes but is not limited to gapped local and quasi-local Hamiltonians). 

Lieb-Robinson bounds are a technical tool for local
lattice models.  In relativistically invariant field theories, the
speed of light is a strict upper bound to the velocity of propagation.
In lattice theories, the LR bounds provide a similar upper
bound by a velocity called the LR velocity, but in contrast
to the relativistic case there can be some exponentially small
``leakage" outside the light-cone in the lattice case.  The
Lieb-Robinson bounds are a way of bounding the leakage outside the
light-cone.  The LR velocity is set by microscopic details
of the Hamiltonian, such as the interaction strength and range.
Combining the LR bounds with the spectral gap enables us to
prove locality of various correlation and response functions. We will call a Hamiltonian a {\em Lieb-Robinson Hamiltonian} if it satisfies LR bounds.  

We work primarily with a single example, but it should be clear that the concept of Galois conjugation can be widely applied to TQFTs.  The essential idea is to retain the particle types and fusion rules of a unitary theory but when one comes to writing down the algebraic form of the $F$-matrices (also called $6j$ symbols), the entries are now Galois conjugated.  A slight complication, which is actually an asset, is that writing an $F$-matrix requires a gauge choice and the most convenient choice may differ before and after Galois conjugation.

Our method is not restricted to Galois conjugated $\textrm{DFib}^{\mathcal{G}}$  and its factors $\textrm{Fib}^{\mathcal{G}}$ and $\overline{\textrm{Fib}^{\mathcal{G}}}$, but can be generalized to infinitely many non-unitary TQFTs, showing that they will not arise as low energy models for a gapped 2D quantum mechanical system with topological order.  

The 2D quantum mechanical systems which can be described by {\em any} type of TQFTs are known as topological phases.  Although this concept is widely noted in the condensed matter physics literature, our introduction is not complete without providing a definition.  Many authors focus on properties (\eg existence of anyonic excitations), but we prefer to give a more fundamental definition since it is this definition which figures into our proof in Sec. \ref{sec:proof}.  We say that a LR Hamiltonian describes a {\em topological phase} (or a phase is {\em topologically ordered}) if and only if its ground state manifold $G$ satisfies the following ``code" property with respect to all spacially local operators $L$: the composition 
\begin{equation}
\label{eq:code}
G\overset{\rm{inc}}\hookrightarrow H \xrightarrow{L} H \xrightarrow{{\rm inc}^\dagger} G
%G \stackrel{\textrm{inc}}{\hookrightarrow} H \stackrel{\textrm{inc}^{\dagger}}{\hookrightarrow} G
\end{equation}
is a multiplication by some scalar $s(L)$ (possibly $s(L)=0$).  $L$ local means $L$ acts only on sites in a sufficiently small radius.  This definition first appeared in Ref.~\onlinecite{Freedman01} (see definition $3.6$ there), conceptualizing the earlier formulation of Ref.~\onlinecite{Gottesman}.  Recently Bravyi {\em et al.} \cite{BHM} have called this axiom ``TQO-1" and advocated an additional requirement that they called ``TQO-2", which enforces a consistency between local and global ground states.  While technically necessary, we know of no realistic case where the second axiom would be required and so have not included it in the framework of this paper.\cite{FootNoteBravyi}

\section{Levin-Wen model and its Galois conjugates}

\subsection{The Levin-Wen model}

Topological quantum field theories are highly constrained mathematical constructs \cite{W,W06, Turaev} designed to capture the low energy physics of topologically ordered systems.  Chern-Simons theory \cite{Witten89} generates most of the known examples; the simplest of these, all chiral, being based on a Lie group and level $k$, $G_k$.  
Starting from a set of particles and fusion rules, there is a standard construction -- called the ``quantum double" or ``Drinfeld center" --
which produces an achiral TQFT. 
Such quantum doubles were introduced in the physics literature by Levin and Wen \cite{LW} in the form of ``string-net" Hamiltonians.
%Levin-Wen models \cite{LW} are ``string-net" Hamiltonians for such doubled theories.
% obtained by taking the mathematical constraints identified in Tuarev-Viro \cite{TV} and writing them as local (commuting) projectors.  
If, for instance, we take the particles and fusion rules from the chiral Fib TQFT, see Eq.\eqref{eq:FusionRules} below, 
and use these to label string-nets on surfaces, a ``larger" TQFT DFib $\cong$ $\rm Fib$ $\otimes\,\,\overline{\rm{Fib}}$ (with more particle types) is obtained.
%there is a standard way of constructing an even larger TQFT, \ie one with more particles.
%If the particle types (``labels") of one TQFT (typically of $G_k$) are used to label string-nets on surfaces, a ``larger" TQFT (with more particle types) is obtained called variously: .  

The Levin-Wen model thus is a microscopic spin Hamiltonian implementing doubled topological theories.
Originally, it was defined \cite{LW} on a honeycomb lattice, but its extension to any trivalent graph is straight-forward. Given a lattice graph and an anyonic theory, the model's Hilbert space is spanned by all labelings of graph edges with the theory's particle types which are consistent with a set of constraints given by the theory, the so-called fusion rules. As a simple example we first consider the Fibonacci theory Fib, where there are only two particle types, namely a trivial particle $\bone$ and the Fibonacci anyon $\btau$. Two particles can combine according to the fusion rules
\begin{align}
\bone \times \bone &= \bone & \bone \times \btau &= \btau & \btau \times \btau = \bone + \btau \ .
\label{eq:FusionRules}
\end{align}
In the Levin-Wen model implementing the doubled Fibonacci theory DFib, this amounts to the constraint that of the three edges meeting in any single vertex never only one can carry a $\btau$ label. This Hilbert space can either be understood as that of an anyonic quantum liquid enclosing the lattice links\cite{Gils09} or alternatively as the the ground states of a spin model (by identifying particle types with spin directions) with a peculiar three-spin interaction enforcing the vertex constraint.

\begin{figure}[t]
	\begin{center}
	\includegraphics[width=.3 \columnwidth]{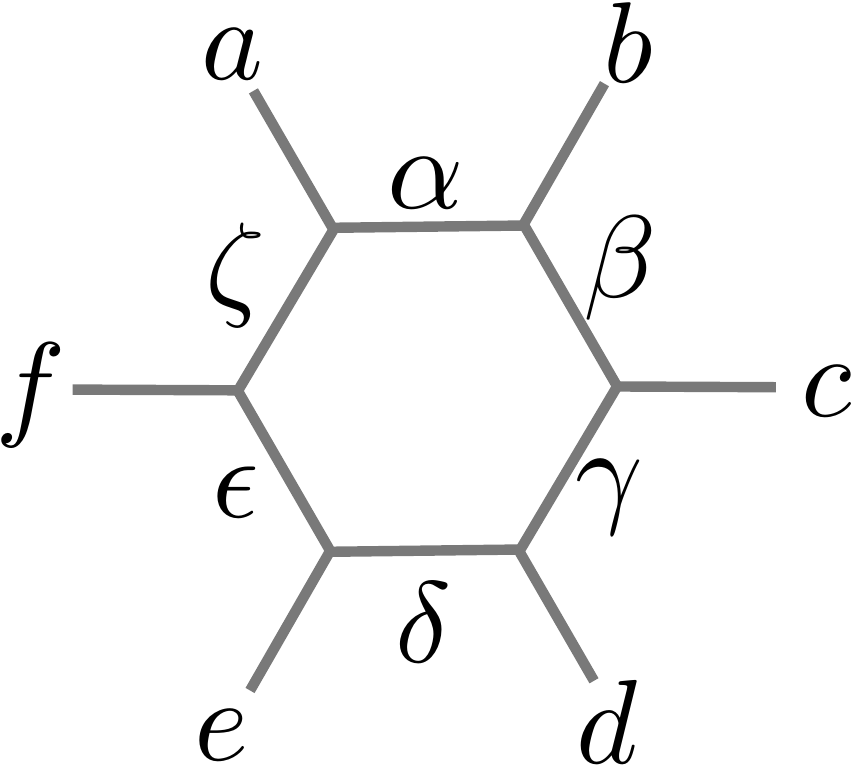}
	\caption{Edge labeling for a plaquette of the honeycomb lattice.}
	\label{fig:honey_plaq}
	\end{center}
\end{figure}

Within these states, the Hamiltonian
\begin{align}
H^\text{LW} &= J_p \sum_{\text{plaquettes } p} \delta_{\phi(p),\btau}
\end{align}
is a projector onto the $\btau$-flux state of a plaquette $p$ thus favoring the trivial flux $\phi(p) = \bone$ through each plaquette. 
The action of this operator on an element of the basis where the edges belonging to plaquette $p$ carry labels $\alpha, \dots, \zeta, a, \dots, f$ as displayed in Fig.\ \ref{fig:honey_plaq} results in a superposition of states where the inner edges of the plaquette carry new labels $\alpha', \dots, \zeta'$ whereas all other edges remain unchanged. Any of the labels takes one of the values $\{ \bone,\btau \}$. The matrix elements between these basis states read explicitly (see Refs.~\onlinecite{LW} and \onlinecite{Gils09} for a detailed derivation)
\begin{align}
\delta_{\phi(p),\btau} =
\bone - 
\sum_{s}  %\sum_{a,\dots,f} \sum_{\substack{\alpha,\dots,\zeta \\ \alpha',\dots,\zeta'}}
&
\frac{d_{s}}{D^{2}}
\left( F^{\alpha' s \zeta}_{a} \right)_{\alpha}^{\zeta'}
\left( F^{\beta' s \alpha}_{b} \right)_{\beta}^{\alpha'}
\left( F^{\gamma' s \beta}_{c} \right)_{\gamma}^{\beta'}
\nonumber\\
\times &
\left( F^{\delta' s \gamma}_{d} \right)_{\delta}^{\gamma'}
\left( F^{\epsilon' s \delta}_{d} \right)_{\epsilon}^{\delta'}
\left( F^{\zeta' s \epsilon}_{d} \right)_{\zeta}^{\epsilon'}
,
\end{align}
where $d_s$ denotes the quantum dimension of particle type $s$, \ie
$d_\bone = 1$ and $d_\btau = \phi \equiv (1+\sqrt{5})/2$, the golden ratio and
 $D$ the total quantum dimension,
$D = \sqrt{d_\bone^2+d_\btau^2} = \sqrt{2+\phi}$ for Fibonacci
anyons. 
For different plaquette geometries this operator has an analogous form with one $F$-symbol for each edge of the plaquette.

\begin{figure}[b]
	\begin{center}
	\includegraphics[width=.8 \columnwidth]{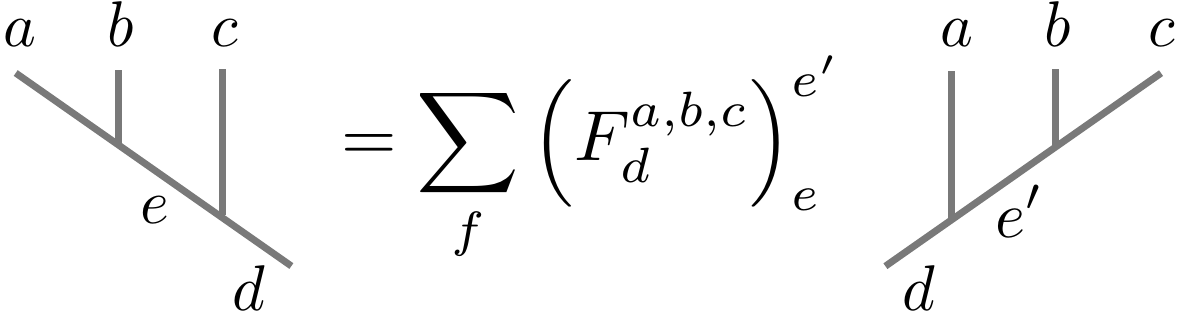}
	\caption{The $F$-symbol.}
	\label{fig:fsymbol}
	\end{center}
\end{figure}

The $F$-symbol, which can be thought of as a generalized $6j$ symbol, describes local basis transformations in a fusion tree as shown in Fig. \ref{fig:fsymbol} and is a defining property of the anyonic theory. For Fibonacci anyons, this transformation is trivial except for the case when all four outer legs of the subgraph that is to be transformed carry the $\btau$ label. Then we have the unitary $2 \times 2$ matrix
\begin{equation}
\label{eq:fsymfib}
   F^{\tau\tau\tau}_\tau
         =  \begin{pmatrix}   \phi^{-1} & c{\phi}^{-1/2} \\
                                     c^{-1}{\phi}^{-1/2} & -\phi^{-1}
          \end{pmatrix} \,.
\end{equation}
where any choice of the gauge $c$ satisfies the pentagon equations for the Fibonacci fusion rules. Choosing the gauge $\c=1$  gives the usual unitary $F$-matrix for the Fibonacci theory, which we refer to as the {\em symmetric} normalization.  From an algebraic point of view the natural gauge choice is $\c=\phi^{5/2}$ which leads to
\begin{equation}
\label{eq:fnatural}
   F^{\tau\tau\tau}_\tau
        =  \begin{pmatrix}   \phi-1 & \phi+1 \\
                                     2\phi-3 & 1-\phi\\
          \end{pmatrix} \ ,
\end{equation}
where no square roots of $\phi$ appear. We refer to this choice as the {\em algebraic} normalization and define $\lambda=c/\phi^{5/2}$. We remark that both normalizations give the same spectra for our models, since the corresponding Hamiltonians are conjugate by a diagonal fugacity change matrix.

The Levin-Wen model can can be solved exactly since all the plaquette terms commute.\cite{LW} As a sum of projectors it counts the number of plaquettes penetrated by a nontrivial $\btau$-flux and the spectrum hence consists of states at all non-negative integer multiples of $J_p$,  corresponding to the number of  nontrivial plaquette fluxes. 

The ground states of the model correspond to all states with no plaquette fluxes, corresponding to the ground states of the topological liquid on a doubled surface around the lattice. With periodic boundary conditions in both directions this surface is a doubled torus with four degenerate ground states.

%, that can be most easily understood in the picture of a thickened lattice. Here plaquettes can be viewed as holes punching the so-obtained surface. If a plaquette is penetrated by a trivial flux only, the hole can be closed continuously. In the ground states of the Levin-Wen model thus all holes can be closed, leaving a single cylinder in the quasi-one-dimensional case and two disconnected sheets in the two-dimensional geometry. Now the cylinder can either carry a trivial or a $\btau$ flux in its interior, which is why this model has two topologically distinct ground states. The two-dimensional case can support a flux in two directions leading to a fourfold ground state degeneracy.
%

\subsection{The doubled Yang-Lee model}

\begin{figure}[t]
\begin{center}
  \includegraphics[width=.9\columnwidth]{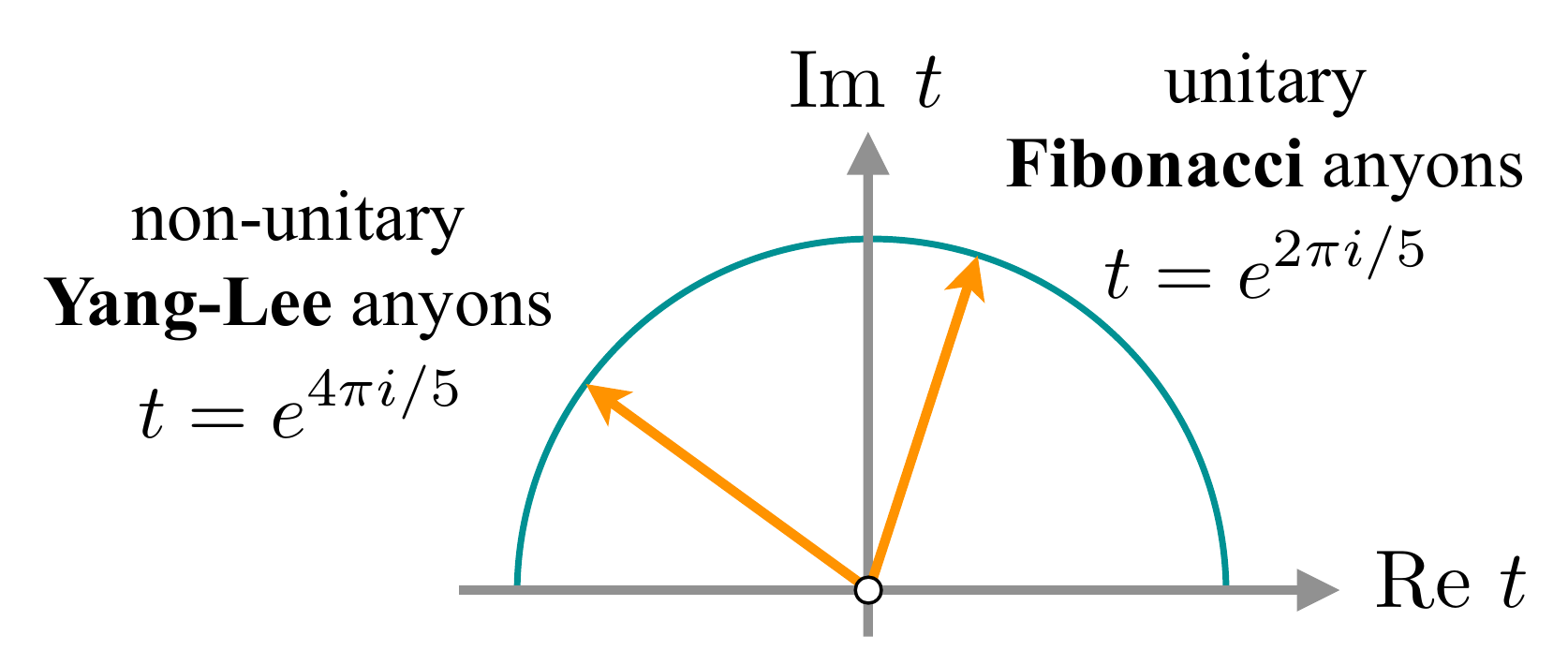}
\end{center}
\caption{(color online) The $t$-deformation parameters of Fibonacci and Yang-Lee anyons correspond 
              to different primitive roots of unity.}
\label{Fig:YangLeeAnyons}
\end{figure}

Now we turn to a theory of non-unitary non-abelian anyons which are closely related to the Fibonacci ones by Galois conjugation. We start by noting that Fib is only one particular sub-theory out of a discrete set of $su(2)_k$ (for finite $k$) anyonic theories, specifically the integral spin half of the unitary $su(2)_3$ theory. The $su(2)_k$ theories are certain deformations of $SU(2)$ characterized by the truncation level $k$, which defines the particle types in the theory, and additionally a deformation parameter $q$ determining the precise values of the $F$-symbol. For $k=3$ two sub-theories characterized by the roots of unity $t=e^{2 \pi i/5}$ and $t'=e^{4 \pi i/5}$ are Galois conjugates of each other (see Fig.~\ref{Fig:YangLeeAnyons}). 

Now, the former value for $q$ produces the Fibonacci theory as described above whereas the latter leads to the non-unitary $F$ matrix
\begin{equation}
\label{eq:falgyl}
   F^{\tau\tau\tau}_\tau
        =  \begin{pmatrix}   d-1 & d+1 \\
                                     2d-3 & 1-d\\
          \end{pmatrix},
\end{equation}
in the algebraic normalization and
\begin{equation}
\label{eq:fsymyl}
   F^{\tau\tau\tau}_\tau
         =  \begin{pmatrix}   -\phi &i{\phi}^{1/2} \\
                                     i{\phi}^{1/2} & \phi
          \end{pmatrix} \,.
\end{equation}
in the symmetric normalization. These are just the $F$-matrices of the DFib theory with $\phi=-t^{1/2}-t^{-1/2}$ replaced by $d=-1/\phi=-t'^{1/2}-t'^{-1/2}$. Here we choose the fourth roots $t^{1/4}= ie^{\pi i/10}$ and $t'^{1/4}= ie^{-3\pi i/10}$, which will be needed to specify the Galois conjugation of the full theory below.

We remark on the choice of algebraic normalization for the $F$ matrix. For the Yang-Lee theory no choice of $c$ would make $F$ unitary---a manifestation of the non-unitarity.  While there would be no topological invariant positive definite Hermitian products on all ground state manifolds in Yang-Lee theory,  there is always a topological invariant Hermitian product with possibly mixed signatures.  The topological invariant inner product is Hermitian because the partition function under time reversal in a (2+1)-topological theory is Hermitian conjugated.  The $(1,2)$-entry of the above $F$-matrix is the theta symbol (the norm of a fusion basis in a fusion space) multiplied by $d^{-2}$, hence it should be a real number.  The above choice of $F$-matrix for Fibonacci case is pleasant when we work on number theory related problems.  In particular, one
notices that in this case we obtain the Galois conjugate by replacing all occurrences of the golden ratio $\phi$ by $d=-1/\phi$, which is the second solution of the quadratic equation $x^2=1+x$.

As Galois conjugation does not change the theory's algebraic structure the doubled Yang-Lee (DYL) Levin-Wen model using the $F$ matrix of Eq. (\ref{eq:fsymyl}), can be solved in exactly the same way as its DFib counterpart. In particular, it has exactly the same spectrum whose eigenvalues count the number of plaquettes penetrated by a non-trivial flux and the same ground state degeneracies. The DYL model also retains the topological protection of the ground state degeneracy against local perturbations.

\section{Hermitian model from non-unitary theory}

\subsection{Constructing Hermitian models}

While the non-Hermitian DYL model features a generalized stable topological phase and a generalized code property, discussed in more detail below, an immediately arising question is whether this phase can also be realized in a Hermitian model. 
There are multiple ways to obtain a Hermitian model %from a non-Hermitian parent model 
that has the same ground states as the non-Hermitian parent model. However, as we will see in the following the question whether the 
topological nature of the ground state remains is a more subtle one.

The simplest Hermitian model $H^\dag H$ is obtained by squaring the non-Hermitian parent Hamiltonian $H$. This model has the same right ground-state eigenvectors as the original model. Alternatively, $H H^\dag $ has the same left ground-state eigenvectors.
The simplicity of this approach comes at the cost of a Hamiltonian which is highly non-local.
To avoid non-local terms, we can take an alternative route and individually square each plaquette term of
$H_p=\delta^\text{DYL}_{\phi(p),\btau}$, arriving at the Hamiltonian $\sum_p H_p^\dag H_p$ or $\sum_p H_p H_p^\dag$. 
Since each plaquette term annihilates the ground state, squaring them in this way also annihilate the  (right/left) ground state eigenvectors.
Finally, we can replace the non-Hermitian plaquette operator $H_p$ with a projector onto the 
complement of the operator's kernel. More specifically, we diagonalize the plaquette operator and use its orthogonalized 
right eigenvectors $\ket{ 0^{(r)}_i }$ belonging to the eigenvalue $0$ to define a projector
\begin{align}
\mathcal{P}_{p} & = 1 - \sum_{i} \ket{ 0^{(r)}_i } \bra{ 0^{(r)}_i } .
\end{align}
The sum of these projectors is then used to define the Hermitian Hamiltonian
\begin{equation}
H^\text{herm} = J_p \sum_p \mathcal{P}_{p}
\; .
\end{equation}
It turns out that all three approaches result in the same qualitative behavior 
-- a loss of the code property and the associated stable topological order -- and we will limit our discussion to the last approach.

\subsection{Loss of the code property}

We find that the non-Hermitian models are stable against local perturbations, and they satisfy a generalized code property. Keeping in mind that a non-Hermitian matrix has left and right eigenvectors, which in general are not identical, a local operator acts as a scalar multiple of an identity operator connecting the left and right ground state subspaces:
\begin{equation}
\bra{0_i^{(l)}}L \ket{0_j^{(r)}} = \lambda(L) \delta_{ij} 	\;.
\end{equation}

Independent of the way we derive a Hermitian model from the parent DYL model, we find that the code property is lost for the Hermitian models: when constructing a Hermitian model, one inevitably has to decide wether to preserve left or right ground states. 
The code property for the Hermitian model would require expectation values of local operators of the form
\begin{equation}
\bra{0_i^{(r)}} L \ket{0_j^{(r)}}  \qquad {\rm and }\qquad \bra{0_i^{(l)}} L \ket{0_j^{(l)}} 
\end{equation}
to again be multiples of the identity. In general, this usual code property will not be satisfied, as one can see, for example, by calculating the matrix elements of a local observable such as a string tension. 
%This is further substantiated by our numerics on the DYL model (discussed below in detail), which show that the matrix between left and right ground states is indeed just a scalar multiple of the identity matrix whereas the matrix between right and right ground states is non-trivial with different eigenvalues. 
Perturbing any Hermitian Hamiltonian which has the (right or left) DYL ground states with an arbitrary small string tension will hence immediately lead to a splitting of the ground-state degeneracy, as we will discuss below.

\subsection{Absence of topological order}

In this section we probe whether topological order survives the construction of a Hermitian model by numerically diagonalizing the models on different
lattice geometries, the honeycomb lattice of the original Levein-Wen construction \cite{LW} and the two-leg ladder geometry of Ref.~\onlinecite{Gils09}.
We diagonalized systems with up to 24 edges using a dense eigenvalue solver and employed iterative schemes for systems with up to 39 edges: the Lanczos algorithm for Hermitian models and an implicitly restarted Arnoldi method for non-Hermitian models.

\subsubsection{Honeycomb model}

\begin{figure}[t]
\begin{center}
\href{http://archive.comp-phys.org/phys.ethz.ch/gukel/paperdata/galois/honey_gap_L.vtl}{
\includegraphics[width= \columnwidth]{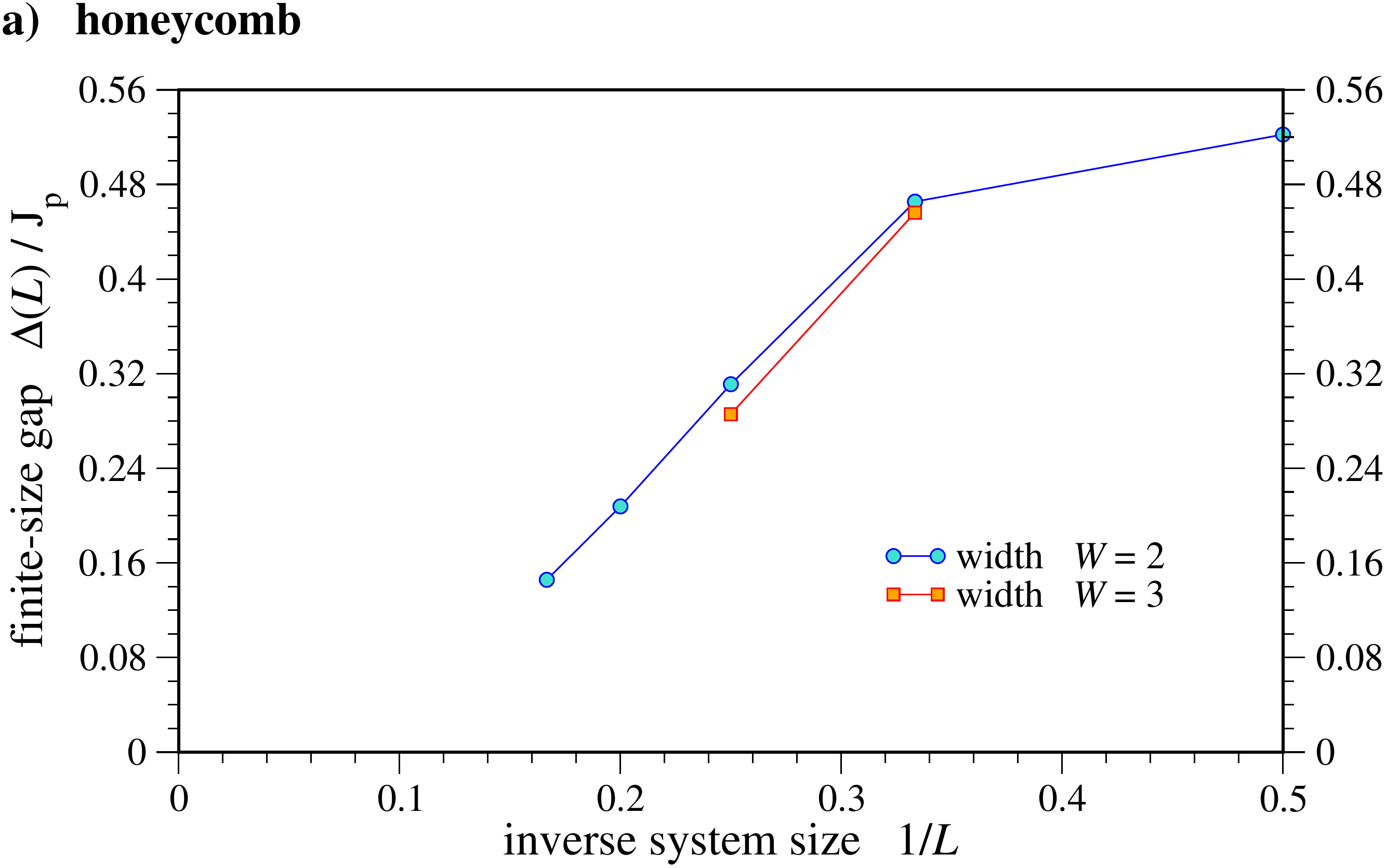}}
\vskip 3mm
\href{http://archive.comp-phys.org/phys.ethz.ch/gukel/paperdata/galois/ladder_gap_L.vtl}{
\includegraphics[width= \columnwidth]{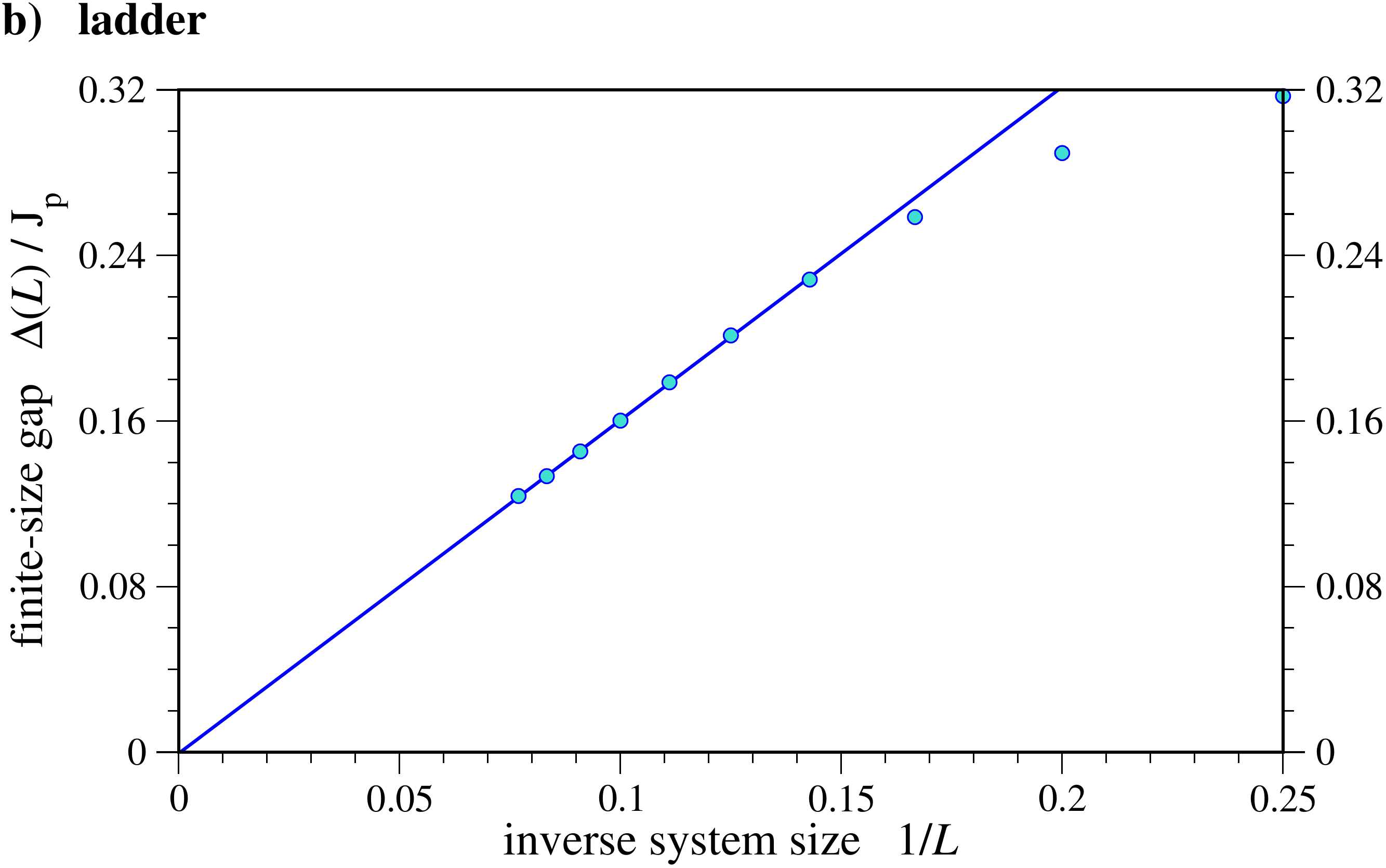}}
\caption{(color online)
	 Scaling of the finite-size gap $\Delta(L)$ (in units of $J_p$) with linear system size for the Hermitian projector model $H^{\rm herm}$
	 on two different lattice geometries: the honeycomb lattice with $L \times W$ plaquettes (top panel) and 2-leg ladder systems
	 of length $L$ (bottom panel).}
\label{fig:gap_scaling}
\end{center}
\end{figure}

Our results on the honeycomb lattice show a clear distinction between the DFib and DYL models on the one hand and the Hermitian model $H^{\rm herm}$ derived from the DYL model on the other hand. While all models feature four degenerate ground states, the former two are gapped, whereas the latter one turns out to be gapless in the thermodynamic limit; see the finite-size extrapolation in Fig.~\ref{fig:gap_scaling}a).
Furthermore, the ground-state degeneracy is easily lifted by a local perturbation, such as a string tension -- in contrast to the stability of the topological phases of the DFib and DYL models.

\subsubsection{Ladder model}

\begin{figure}[t]
	\begin{center}
	\includegraphics[width=.35 \columnwidth]{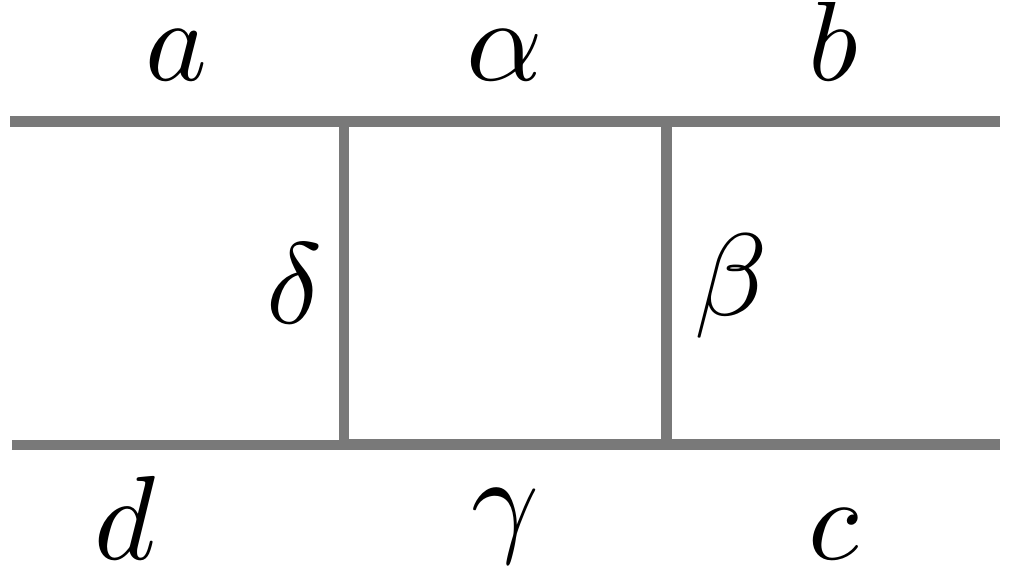}
	\caption{Edge labeling for a plaquette of the ladder lattice.}
	\label{fig:ladder_plaq}
	\end{center}
\end{figure}

Since only small linear dimensions are accessible to exact numerical diagonalization for the honeycomb lattice, we also consider a a quasi-one-dimensional ladder geometry consisting of rectangular plaquettes as shown in Fig.~\ref{fig:ladder_plaq}.
The DFib and DYL models on this ladder geometry were introduced and solved in Refs.~\onlinecite{Gils09} and \onlinecite{YLC}, respectively. Both models feature  topological phases with two (instead of four) degenerate ground states, but are otherwise identical to the respective honeycomb lattice models.

%\subsubsection{Perturbation with a string tension}

\begin{figure}[b]
	\begin{center}
	\href{http://archive.comp-phys.org/phys.ethz.ch/gukel/paperdata/galois/ladder_dyl_gap_theta.vtl}{
	\includegraphics[width=\columnwidth]{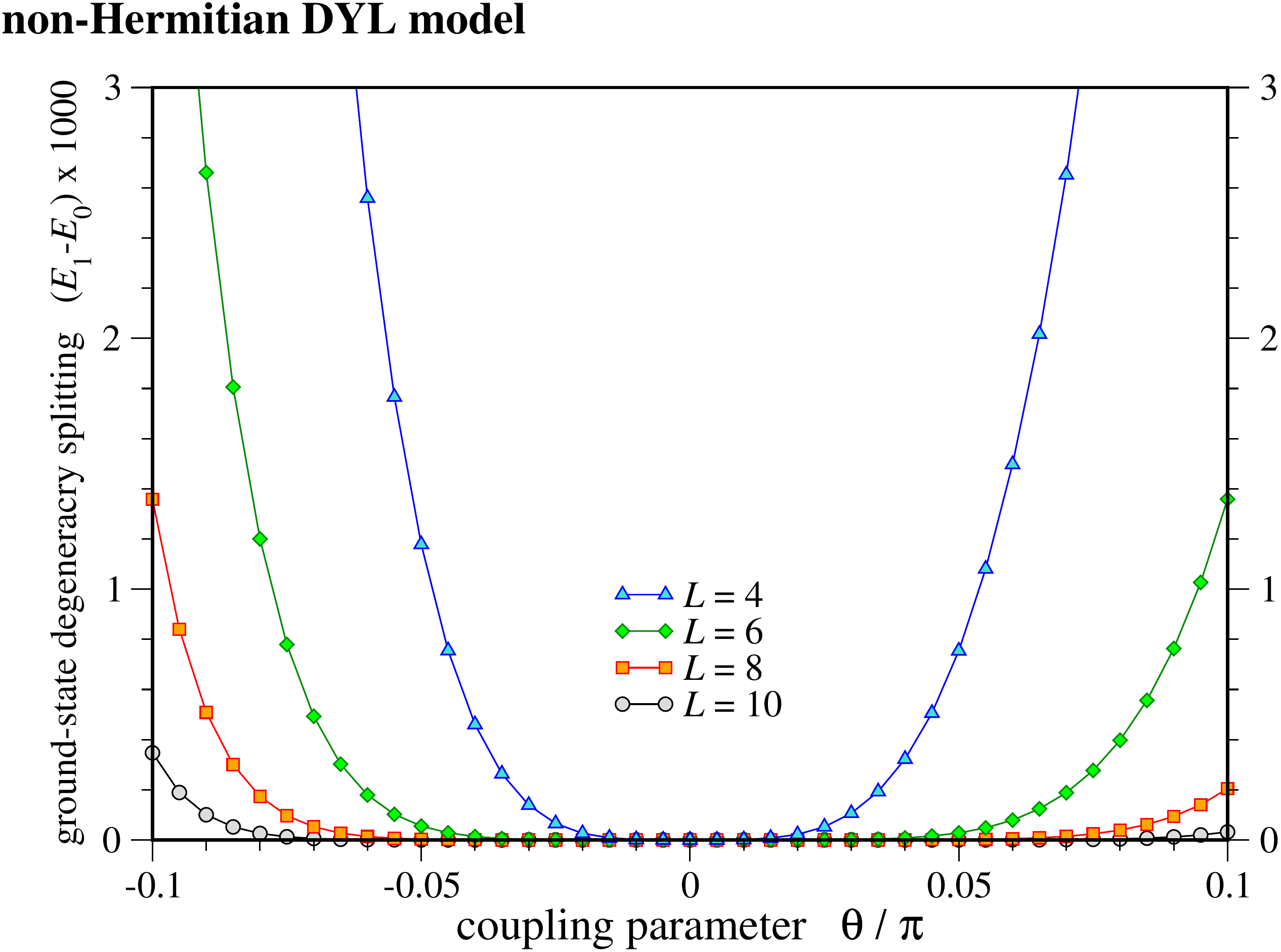}}
	\caption{(color online)
	              Ground-state degeneracy splitting of the non-Hermitian doubled Yang-Lee model
	              when perturbed by a string tension $(\theta \neq 0)$.}
	\label{fig:dyl_deg_splitting}
	\end{center}
\end{figure}

\begin{figure}[t]
	\begin{center}
	\href{http://archive.comp-phys.org/phys.ethz.ch/gukel/paperdata/galois/ladder_gap_theta.vtl}{
	\includegraphics[width=\columnwidth]{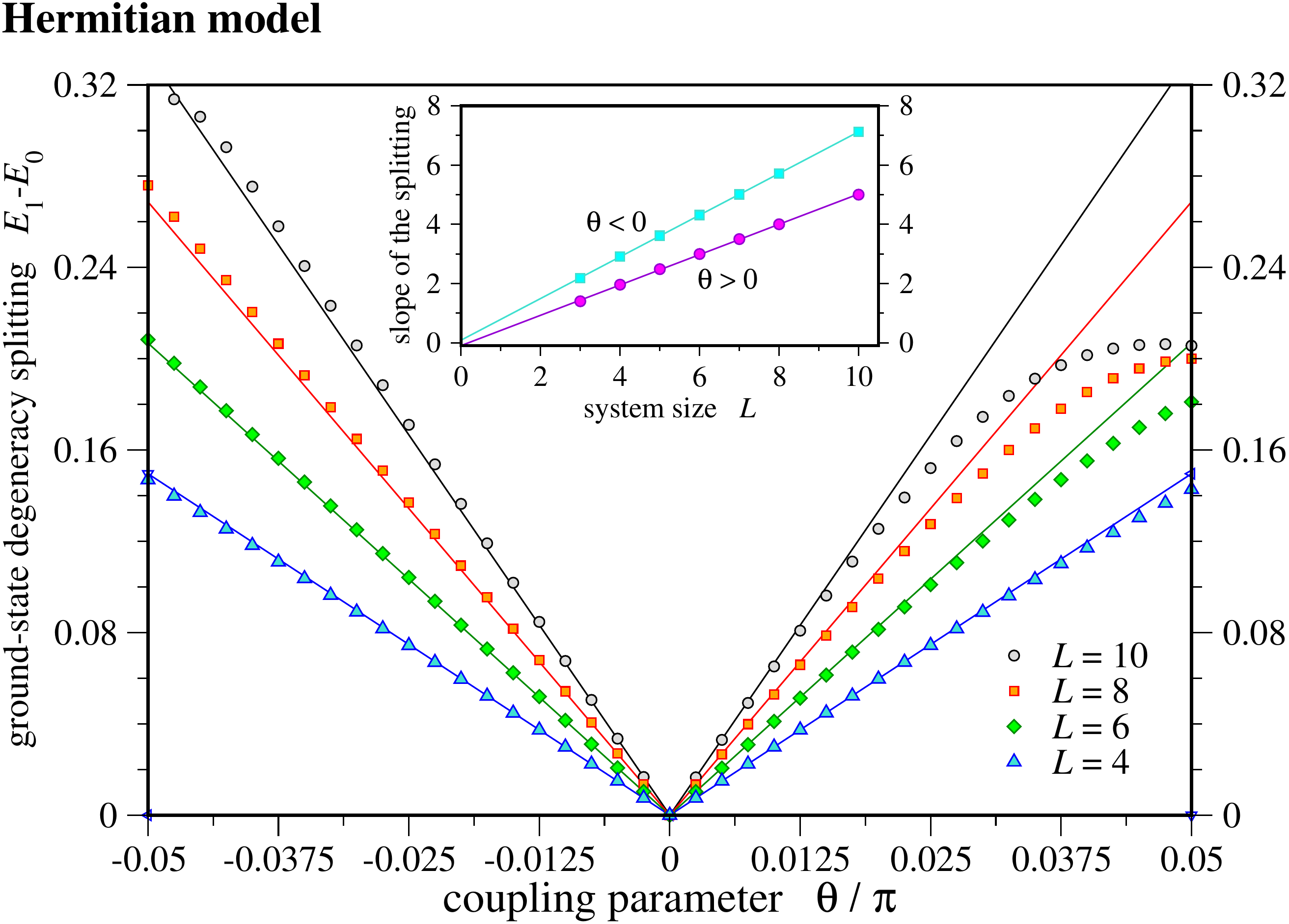}}
	\vskip 3mm
	\href{http://archive.comp-phys.org/phys.ethz.ch/gukel/paperdata/galois/ladder_E_around_theta0.vtl}{
	\includegraphics[width= \columnwidth]{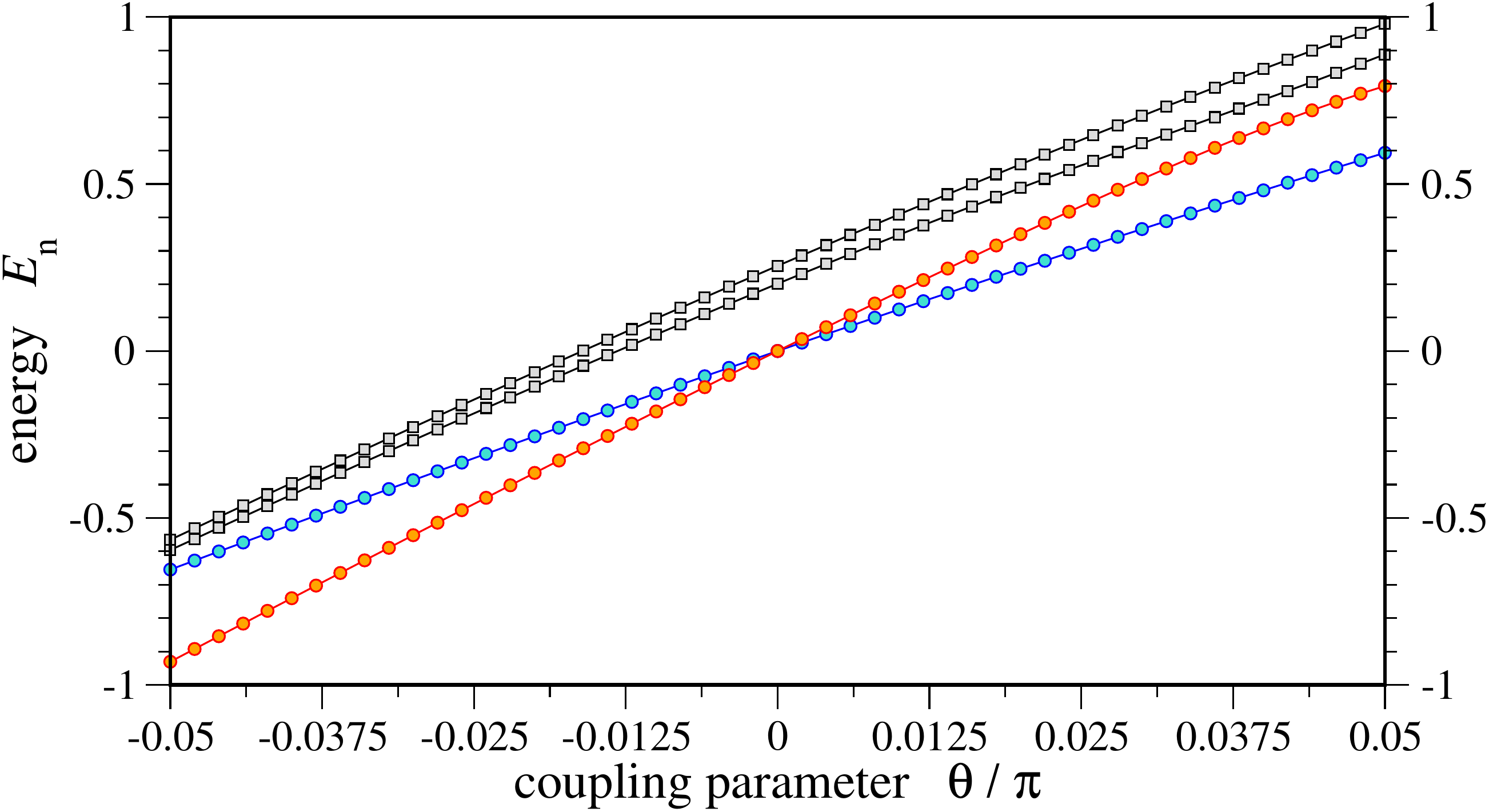}}
	\caption{(color online)
	              Ground-state degeneracy splitting of the Hermitian model $H^{\rm herm}$, the counterpart to the DYL model,
	              when perturbed by a string tension $(\theta \neq 0)$ (top panel).
	             The slope of the splitting around the unperturbed model $(\theta = 0)$ is given in the inset (top panel)
	             for different system sizes $L$.
	             The bottom panel shows the low-energy spectrum, which clearly shows that the degeneracy at $\theta=0$ is 
	             due to a level-crossing.
	 }
	\label{fig:proj_deg_splitting}
	\end{center}
\end{figure}

The quasi-one dimensional geometry allows to numerically diagonalize systems up to linear system size $L=13$. 
The finite-size gap of the Hermitian model $H^{\rm herm}$ is again found to vanish in the thermodynamic limit, showing a linear 
dependence on the inverse system size as shown in Fig.~\ref{fig:gap_scaling}b).
To further demonstrate the fragility of these gapless ground states  against local perturbations we add a string tension~\cite{Gils09} 
\begin{equation}
   H^\text{pert} = J_r \sum_{\text{rungs } r} \delta_{l(r),\btau}
\end{equation}
favoring the trivial label $l(r)=\bone$ on each rung of the ladder. 
We parameterize the couplings of the competing plaquette and rung terms as
\[
     J_r=\sin \theta \quad \quad {\rm and} \quad \quad J_p = \cos \theta \,,
\] 
where $\theta=0$ corresponds to the unperturbed Hamiltonian.
The phase diagrams as a function of $\theta$ have been mapped out for both the DFib model \cite{Gils09} and the DYL model,\cite{YLC}
respectively.

Directly probing the topological order in the DYL model and its Hermitian counterpart we show the lifting of their respective ground-state degeneracies in Figs.~\ref{fig:dyl_deg_splitting} and \ref{fig:proj_deg_splitting}  when including a string tension. 
We find a striking qualitative difference between these two models: For the DYL model the lifting of the ground-state degeneracy is exponentially suppressed with increasing system size -- characteristic of a topological phase. For the Hermitian model, on the other hand, 
we find a splitting of the ground-state degeneracy proportional to $J_r L$. The linear increase with both system size and coupling can be easily understood by the different matrix elements of the string tension term on a single rung for the two degenerate ground-states of the unperturbed model. 
Plotting the low-energy spectrum in Fig.~\ref{fig:proj_deg_splitting} clearly shows that the two-fold degeneracy of the unperturbed Hermitian model arises from a (fine-tuned) level crossing. Similar behavior is found in the honeycomb lattice model (not shown).

%\subsubsection{Evolution from plaquette to edge term}

\begin{figure}[t]
	\begin{center}
	\href{http://archive.comp-phys.org/phys.ethz.ch/gukel/paperdata/galois/ladder_dyl_spectrum_sweep.vtl}{
	\includegraphics[width= \columnwidth]{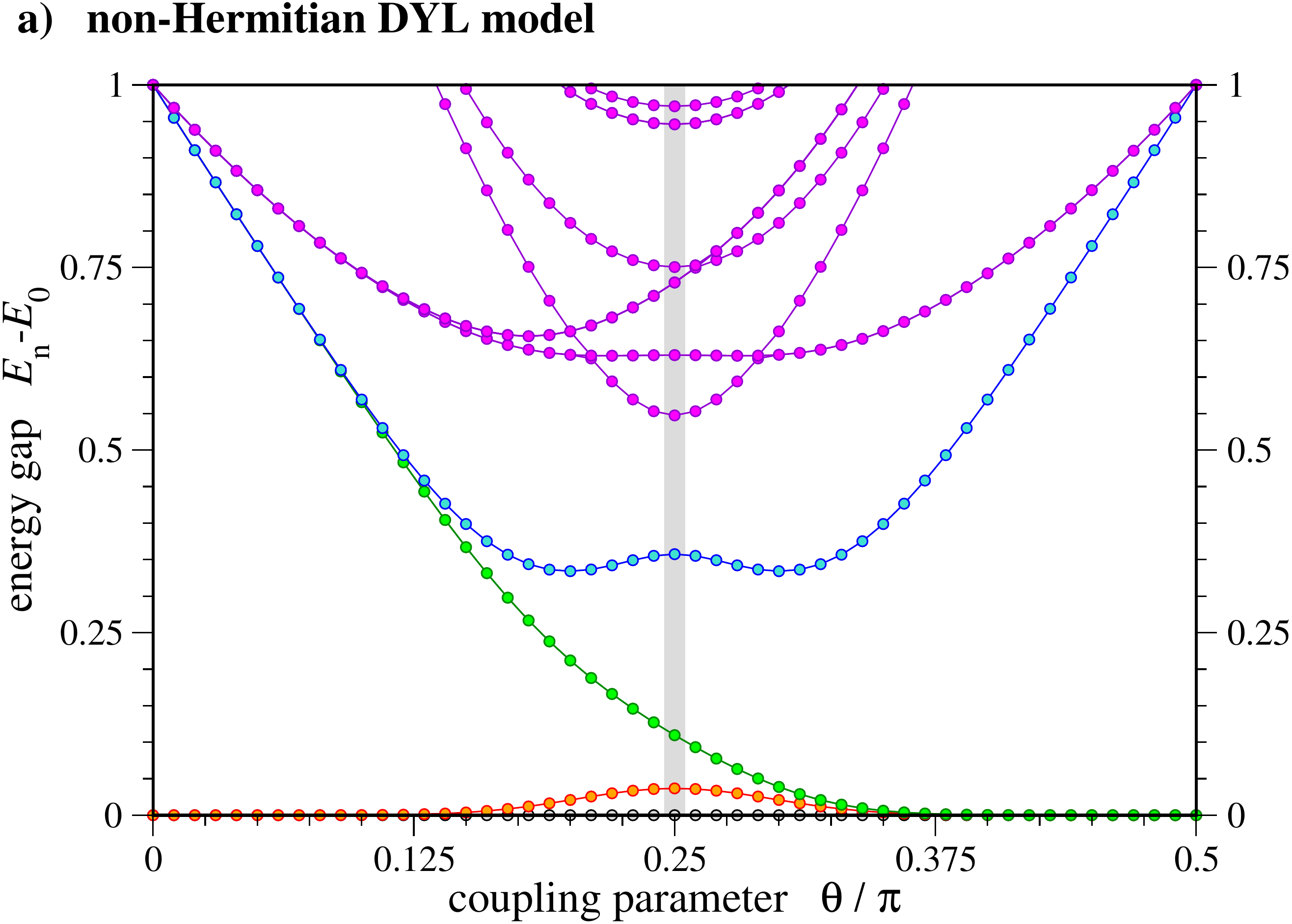}}
	\vskip 3mm
	\href{http://archive.comp-phys.org/phys.ethz.ch/gukel/paperdata/galois/ladder_spectrum_sweep.vtl}{
	\includegraphics[width= \columnwidth]{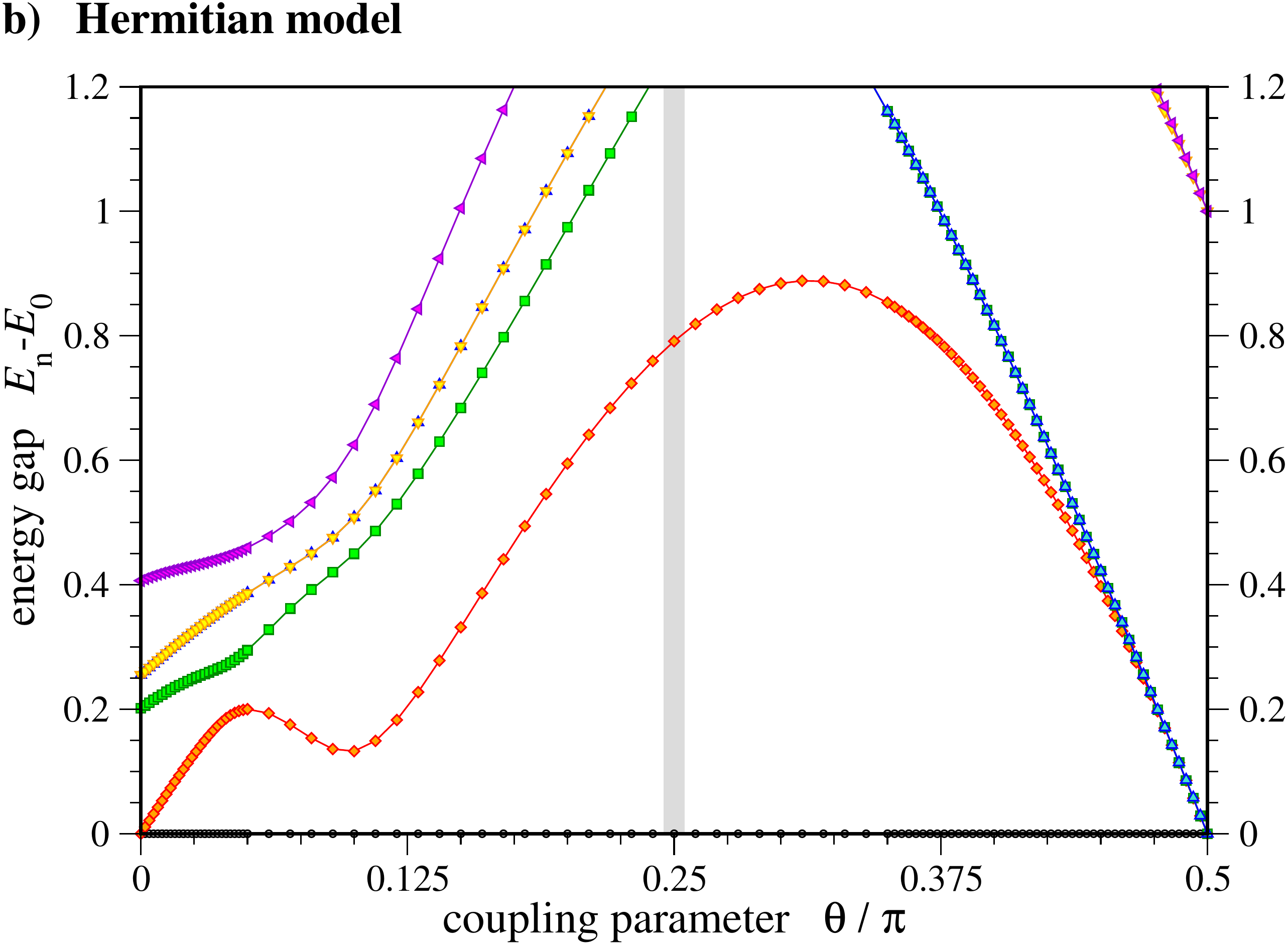}}
	\caption{(color online)
	              The low-energy spectra of the doubled Yang-Lee model (top) its Hermitian counterpart (bottom) 
	              for a wide range of coupling parameters. Data shown is for a ladder of length $L=8$.}
	\label{fig:spectrum_sweeps}
	\end{center}
\end{figure}

Considering the model in a wider range of couplings, as shown in Fig.~\ref{fig:spectrum_sweeps}, further striking differences between the non-Hermitian DYL model and its Hermitian counterpart are revealed: The DYL model exhibits two extended topological phases around $\theta=0$ and $\theta=\pi/2$ (with two and four degenerate ground states, respectively), which are separated by a conformal critical point at precisely $\theta_c = \pi/4$ as discussed extensively in Refs.~\onlinecite{Gils09,YLC}.
In contrast, the Hermitian model $H^{\rm herm}$ exhibits no topological phase anywhere, and the intermediate coupling $\theta=\pi/4$ does not stand out.

\section{Absence of non-unitary topological phases in unitary models}
\label{sec:proof}

So far, we have considered a specific set of Hermitian models constructed to have the same ground states as a non-Hermitian parent model and found that they no longer exhibit a topological phase. This raises the question whether this observation points to a deeper principle, which we investigate in this section in rigorous mathematical terms.

\subsection{Galois Conjugate}
\label{Sec:GaloisConjugate}

Let us now lay out the mathematical foundations as clearly as possible.  
The double DFib is isomorphic to a copy of Fib and its time reversal, DFib $\cong$ Fib $\otimes\,\,\overline{\rm{Fib}}$. Thus to Galois conjugate DFib it is sufficient to define $\rm{Fib}^{\mathcal{G}}$, then $\rm DFib^{\mathcal{G}}\cong Fib^{\mathcal{G}}\otimes \overline{Fib^{\mathcal{G}}}$.

A theory such as Fibonacci can be defined using a set of $6j$-symbols $\{F^{ijk}_{lmn}\}$, braiding eigenvalues $\{R^{bc}_{a}\}$ (not always necessary), and some pivotal coefficients $\{\epsilon_i=\pm 1 \}$, where $i,j,k,l,m,n,a,b,c$ are anyon types (see chapter 4 of Ref.~\onlinecite{ZW}).  Because of gauge choices, there are many different sets for the same theory.  If we fix a set of data, then we can define a number field $K$ for a theory as the number field obtained from adjoining all numbers  $\{F^{ijk}_{lmn}\}$ and $\{R^{bc}_{a}\}$ to the rational numbers $\mathbf{Q}$ ( $\{\epsilon_i=\pm 1 \}$ are already in $\mathbf{Q}$).  The automorphisms of the number field $K$ fixing $\mathbf{Q}$ form the Galois group of $K$, denoted as $G_K$.  If $g$ is an element of $G_K$, then by applying $g$ to all data, we get a potentially new theory.  We will call the new theory a Galois conjugate or a Galois twist.  For the Fibonacci theory, the minimal number fields required for the Galois conjugation for both the algebraic normalization and unitary normalization are worked out in Ref.~\onlinecite{FW} and needed below for the discussion of the projectors for code subspace property.  For the algebraic normalization, the number field is the cyclotomic number field $\mathbb{Q}(\xi_{20})$, where $\xi_N=e^{2\pi i/N}$, while for the unitary normalization, the number field is $\mathbb{Q}(\sqrt{\phi},\xi_{20})$.

It is known in general that a theory from quantum groups such as Fibonacci can always be defined within a cyclotomic field $\mathbb{Q}(\xi_{N})$ for some $N$.  For the Jones representation with the algebraic normalization, this is done explicitly by Kuperberg \cite{GK}. To explain this, we digress briefly to some basic quantum topology.

The Jones representation (and polynomial) may be constructed from the Kauffman bracket: \\

\begin*%{figure}[htpb]
\labellist \small\hair 2pt
  \pinlabel $\text{$=-t^{1/4}$}$ at 120 35
  \pinlabel $\text{$-t^{-1/4}$}$ at 280 40
  \pinlabel $\text{$=-t^{1/2}-t^{-1/2}$,}$ at 650 40
  \pinlabel $,$ at 400 33
\endlabellist
\centering
\includegraphics[scale=.3]{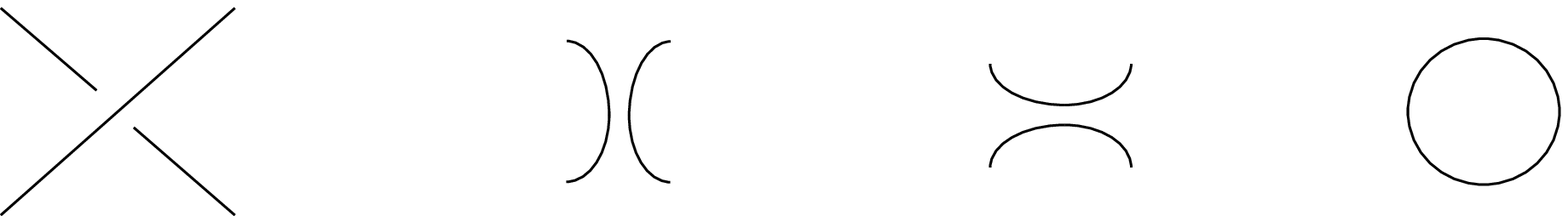}
%\caption{} \label{fig_krelation1}
\end*%{figure}

\noindent $t$ indeterminate. 

The skein space $W(n\cdot{1},0,t)$ is the vector space of formal linear combinations of arc matchings (\ie skeins) of an even number, $n$, of fixed points on the top of a square (the arcs are imbedded in the interior of the square and ``0" means no points marked on the bottom of the square). An $n$-braid $b$ acts on $W$, by gluing $b$ onto the top of the square and resolving crossings by the above rule. The trick is to define each braid generator as $t^{1/4}$ times the geometric crossing. Additionally, each skein is ``even" or ``odd" according to whether a $2$-coloring of the skein complement (starting with the bottom of the square being colored white) has an even or odd number of black regions. All black skeins should be multiplied by a factor of $t^{1/2}$. This results in a basis for $W$ and a rescaling of the action so that this ``$W$-representation" is defined over the field $Q[t]$. The famous quantum representations of Jones at level $k$ will be quotients of the $W$-representation for $t=e^{\frac{2\pi{i}}{k+2}}$. Note that the rescaling cannot affect the density of the projective Jones representation, which will be important shortly.

In the case at hand, Fib, $t=e^{2\pi{i}/5}$, and the Galois conjugate theory $\rm{Fib^{\mathcal{G}}}$ is obtained by replacing $t$ by $t'=e^{4\pi{i}/5}$.

The skein space $W(n\cdot{1},0,t)$ carries a natural bilinear form $\langle{\,,\,}\rangle$ obtained by doubling the square (thought of as a disk) along its boundary and evaluating the union of the two skeins as a scalar using the above Kauffman relations. When $|t|=1$, the form $\langle{\overline{A},B}\rangle$ is Hermitian. If further $t$ is a root of unity, then this form has a singular subspace. $X(n\cdot{1},0,t)$ is, by definition, the finite dimensional Hilbert space obtained by annihilating this kernel. The Hermitian form $\langle{\overline{A},B}\rangle$ is non-singular on $X$ and the braid group $B_n$ acts.

When $t=e^{\frac{2\pi{i}}{k+2}}$, this quotient action is the Jones representation associated to $SU(2)_k$, whose trace leads to the Jones polynomial evaluated at $t$. For $t=e^{2\pi{i}/{(k+2)}}$, the Hermitian form $\langle{\overline{A},B}\rangle$ is positive definite. For other roots of unity, $\langle{\overline{A},B}\rangle$ may be of mixed signs $(p,q)$, $p\neq{0}$, $q\neq 0$. This happens in particular for $t'=e^{4\pi{i}/5}$ when $n\geq{4}$ as we now check.

Well-established conventions in mathematics and physics lead to two different ways to label the particle types in $SU(2)_k$: one by the spins of the irreps, and the other by the dimensions of irreps minus one.  Unless we speak explicitly of a spin label, as in the next paragraph, the labels of particles in this section are by the dimensions minus one, which are twice of the physical spins.

Note that $X(n\cdot{2},0,e^{2\pi{i}/5})\cong{X(n\cdot{1},0,e^{2\pi{i}/5})}$ as Hilbert spaces via the `` $\widehat{}$ " automorphism of $SU(2)_3$: $\widehat{spin\,0}=spin\,{3/2}$, $\widehat{spin\,{1/2}}=spin\,{1}$, $\widehat{spin\,{1}}=spin\,{1/2}$, $\widehat{spin\,{3/2}}=spin\,{0}$. Similarly, $X(n\cdot{2},0,e^{4\pi{i}/5})\cong{X(n\cdot{1},0,e^{4\pi{i}/5})}$ as Hilbert spaces of mixed sign. The braid group actions (Galois conjugates of the Jones representation) are, of course, also identical.
For $t=e^{4\pi{i}/5}$, the loop value is $d=-e^{2\pi{i}/5}-e^{-2\pi{i}/5}=-{1/{\phi}}$, $\phi=\frac{1+\sqrt{5}}{2}$, the golden ratio.

We use trivalent graphs with the Kauffman vertex normalization: \\ \\\begin{center}
\begin*%{figure}[htpb]
\labellist \small\hair 2pt
  \pinlabel $\text{$1/2$}$ at 140 -10
    \pinlabel $\text{$1/2$}$ at 240 -10
    \pinlabel $\text{$1/2$}$ at 160 60
    \pinlabel $\text{$1/2$}$ at 210 60
    \pinlabel $\text{$1/2$}$ at 140 25
    \pinlabel $\text{$1/2$}$ at 240 25
    \pinlabel $\text{$=$}$ at 100 40
    \pinlabel $\text{$1$}$ at 40 90
    \pinlabel $\text{$1$}$ at -7 0
    \pinlabel $\text{$1$}$ at 80 0
\endlabellist
\centering
\includegraphics[scale=.5]{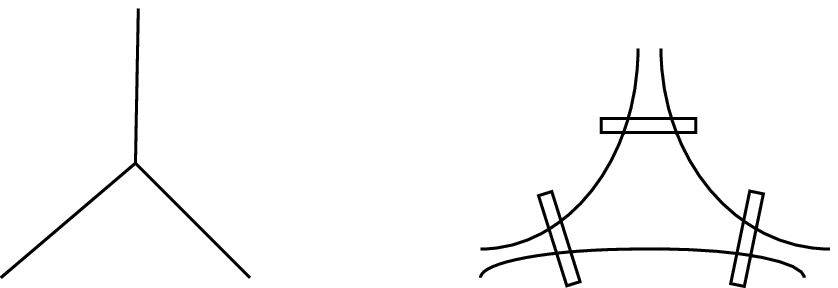}
%\caption{} \label{fig_2t3}
\end*%{figure}
\end{center}

\vspace{.1cm}

\noindent where \begin*%{figure}[htpb]
\labellist \small\hair 2pt
\endlabellist
\centering
\includegraphics[scale=.5]{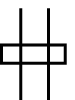}
%\caption{} \label{fig_bar}
\end*%{figure}
is the Jones-Wenzl projector
\begin{center}
\begin*%{figure}[htpb]
\labellist \small\hair 2pt
  \pinlabel $\text{$P_2$}$ at -30 15
    \pinlabel $\text{$-$}$ at 18 15
    \pinlabel $\text{$=$}$ at -12 12
    \pinlabel $\text{$\frac{1}{d}$}$ at 30 15
    \pinlabel $\text{,}$ at 65 8
\endlabellist
\centering
\includegraphics[scale=.5]{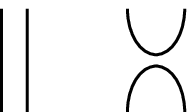}
%\caption{} \label{fig_joneswenzl}
\end*%{figure}
\end{center}
and $1/2$ and $1$ are the spins of the quasi-particles, to write an orthogonal basis for $X(4\cdot{2},0,t)$ (The "$.2$" after $n$ indicates the second, \ie spin $1$, nontrivial particle type).\\
 \begin{center}
\begin*%{figure}[htpb]
\labellist \small\hair 2pt
  \pinlabel $\text{$,$}$ at 180 -5
    \pinlabel $\text{$=$}$ at 380 40
  \pinlabel $\text{$=\frac{1}{d^2},$}$ at 530 40
\endlabellist
\centering
\includegraphics[scale=.3]{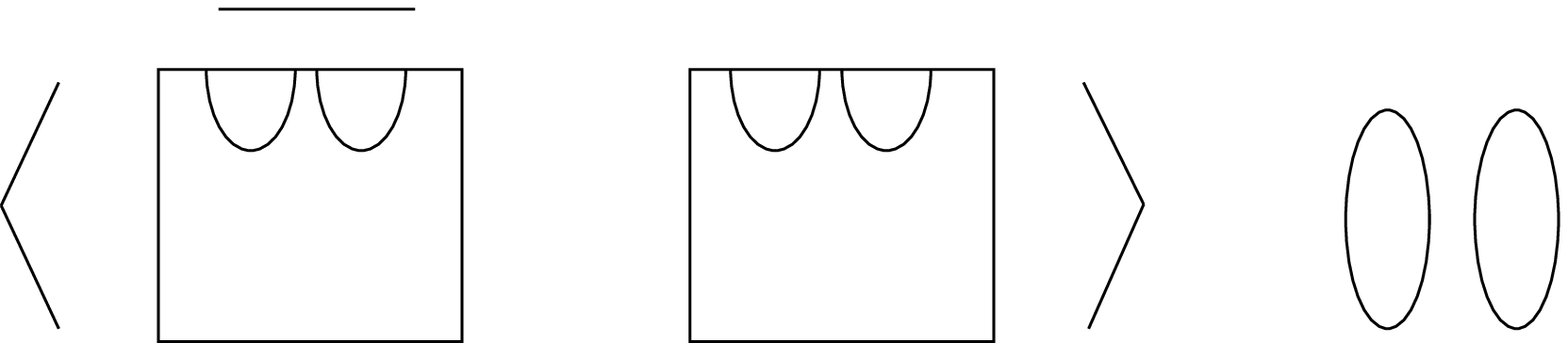}
%\caption{} \label{fig_innerproduct1}
\end*%{figure}
\end{center}
\noindent is a positive number. Now consider an orthogonal basis element
\begin{center}
\begin*%{figure}[htpb]
\labellist \small\hair 2pt
  \pinlabel $.$ at 103 5
\endlabellist
\centering
\includegraphics[scale=.3]{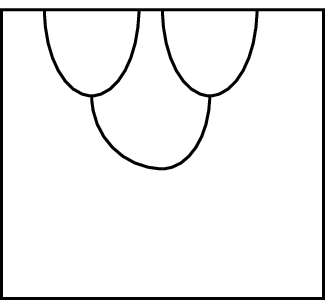}
%\caption{} \label{fig_basiselement}
\end*%{figure}
\end{center}
\noindent We have
\begin{center}
\begin*%{figure}[htpb]
\labellist \small\hair 2pt
  \pinlabel $\text{$,$}$ at 205 -5
    \pinlabel $\text{$=$}$ at 450 50
  \pinlabel $\text{$=0$.}$ at 650 50
\endlabellist
\centering
\includegraphics[scale=.25]{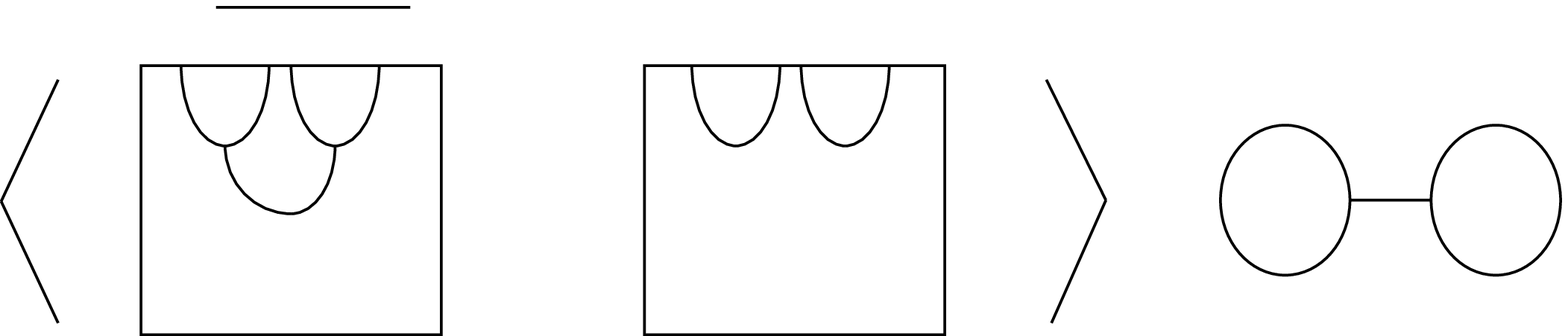}
%\caption{} \label{fig_innerproduct3}
\end*%{figure}
\end{center}
\noindent Its self pairing is negative:
\begin{center}
\begin*%{figure}[htpb]
\labellist \small\hair 2pt
  \pinlabel $\text{$,$}$ at 160 -5
    \pinlabel $\text{$=$}$ at 338 40
  \pinlabel $\text{$=\Big{(}\frac{\theta}{d}\Big{)}^2d,$}$ at 500 45
\endlabellist
\centering
\includegraphics[scale=.3]{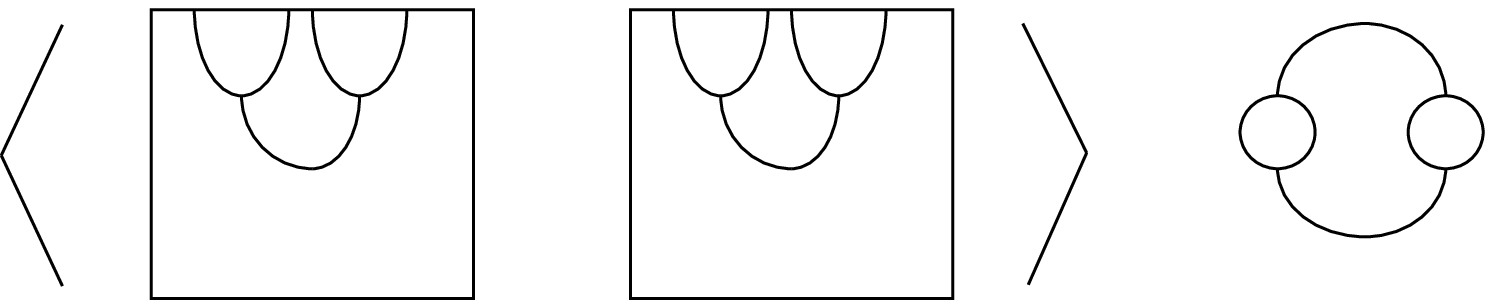}
%\caption{} \label{fig_innerproduct4}
\end*%{figure}
\end{center}
\noindent a negative number where $d$ is the loop value $\frac{-1}{\phi}$ and the $\theta$-symbol $=-\phi$ (simplified from the formula in Figure 19 in Ref.~\onlinecite{LF}). It follows that for $n\geq{4}$ the Hermitian structure on $X(n\cdot{2},0,e^{4\pi{i}/5})\cong{X(n\cdot{1},0,e^{4\pi{i}/5})}$ has mixed signs. The corresponding braid representations for $\rm{DFib^{\mathcal{G}}}$ also have mixed signs when $n\geq{4}$.

The doubled Fibonacci theory DFib, as with all topological phases, has the code property: the composition $G\overset{\rm{inc}}\hookrightarrow H \xrightarrow{L} H \xrightarrow{P} G$ is multiplication by some scalar $\lambda_L \in \mathbf{C}$ whenever $L$ is a (sufficiently) local operator ($P =$ $\rm{inc}^{\dagger}$ is the Hermitian orthogonal projection to the Levin-Wen ground state $G$, see Ref.~\onlinecite{LW}). In general $P$ is Hermitian, but for Fib $P$ is actually real symmetric.

In the Levin-Wen Hamiltonian scheme, there are two kinds of terms: the vertex type $A_v$ for each vertex $v$, and a plaquette type $B_p$ for each plaquette $p$.  The code space $G$ is the common eigenspaces of all local operators $\{A_v\}$ and $\{B_p\}$.  With the standard choices of basis, the vertex terms $A_v$ are matrices with entries $0$'s and $1$'s, while the plaquette terms $B_p$ are matrices with entries given by products of $6j$-symbols.  The algebraic constraints defining the code subspace transform under a Galois conjugation.  Hence
solving the Galois conjugates of the constraints defining $G$, we obtain $G^{\mathcal{G}}$ and replace the matrix $P$ with its Galois conjugate $P^{\mathcal{G}}$. Clearly $G^{\mathcal{G}}\xrightarrow{\rm{inc}} H \xrightarrow{L} H \xrightarrow{P^{\mathcal{G}}} G^{\mathcal{G}}$ is multiplication by $(\lambda_{L^{(\mathcal{G}^{-1})}})^{\mathcal{G}}$ since $L$ is local if and only if $L^{(\mathcal{G}^{-1})}$ is local. Thus $G^{\mathcal{G}}$ retains a ``code" property but with respect to a non-Hermitian projector $P^{\mathcal{G}}$.  In the symmetric normalization of Yang-Lee, $P^{\mathcal{G}}$ is complex symmetric with eigenvalues 0 and 1 (since $(P^{\mathcal{G}})^{2} = P^{\mathcal{G}}$, however $(P^{\mathcal{G}})^{\dagger} = \bar{P^{\mathcal{G}}} \not= P^{\mathcal{G}}$).

We close this paragraph by noting that our proof uses the mixed signatures in the Jones braid group representations as shown above, and as such applies only to the algebraic normalization. For the symmetric normalization of the DYL theory, the Jones representation spaces have either positive or negative definite inner products, but the mixed sign in the algebraic normalization will be sufficient to prove our theorem for any choice of normalization.

\subsection{Lieb-Robinson bounds and local unitary evolution of a ground state under changes in the Hamiltonian}

Our proof of absence below will be based on a contradiction of the above result for the Galois conjugated theory with local unitary evolution of a ground state under local changes in a Hermitian Hamiltonian. This local unitary evolution can be proven for all Hermitian Hamiltonians that satisfy Lieb-Robinson bounds.

Lieb-Robinson bounds are a mathematical way of expressing the physical fact that in local lattice Hamiltonians there is some
upper bound to the velocity of excitations.
These bounds can be proven for a wide-range of Hamiltonians, including what would be
colloquiually referred to as ``Hamiltonians with finite-range interactions" or ``Hamiltonians with exponentially decaying interactions".

For a precise statement of conditions under which Lieb-Robinson bounds can be proven, we follow Ref. \onlinecite{hastingskoma} where one sufficient condition
is given as follows (see also Ref. \onlinecite{ns}).  We consider lattice Hamiltonians, and use $i,j,...$ to label sites of the lattice, with some
metric ${\rm dist}(i,j)$ on the lattice.  We use $X,Y,Z,...$ to label
sets of sites of the lattice.
Let the Hamiltonian $H$ be written as
\be
H=\sum_Z H_Z,
\ee
where the operators $H_Z$ are supported on sets $Z$ (an operator is said to be supported on a set $Z$ if it can be written as a tensor product of an operator on the degrees of freedom on set $Z$ with an identity operator on the remaining degrees of freedom).
Assume that the following condition holds for all sites $i$, 
\begin{equation}
\label{Hdecay}
\sum_{X\ni i}
\Vert H_X\Vert|X|\exp[{\mu\>{\rm diam}(X)}]\le s<\infty,
\end{equation}
for some positive constants $\mu,s$, where ${\rm diam}(X)$ denotes the diameter of set $X$, and $|X|$ denotes the cardinality of
$X$, and $\Vert  H_X \Vert$ denotes the operator norm.

Then,\cite{hastingskoma} Eq.~(\ref{Hdecay}) implies the following Lieb-Robinson bound for Hermitian Hamiltonians.
For any operator  $O$, we use $O(t)$ to denote the Heisenberg time evolution of the operator: $O(t)=\exp[i H t] O \exp[-i H t]$.
Let $A_X,B_Y$ be operators supported on sets
$X,Y$, respectively.
Then there is a constant $v_{LR}$ depending only on $s,\mu$ such that for
$t$ real with $|t|\leq {\rm dist}(X,Y)/v_{LR}$, we have
\be
\label{lrbound}
\Vert [A_X(t),B_Y]\Vert \leq
\frac{v_{LR} |t|}{l} g(l) |X| \Vert A_X \Vert \Vert B_Y \Vert,
\ee
where $l={\rm dist}(X,Y)$ and $g(l)$ decays exponentially in $l$.
Given that $A_X(t)$ has small commutator with all operators $B_Y$ with sufficiently large distance ${\rm dist}(X,Y)$, this implies\cite{bhv}
that the operator
$A_X(t)$ can be approximated by an operator $A_X^l(t)$ which is supported on the
set of sites within distance $l=v_{LR} |t|$ of the set $X$ up to an error in operator norm which is bounded by
$\frac{v_{LR} |t|}{l} g(l) |X| \Vert A_X \Vert$.

\begin{definition}
We say that a Hamiltonian is a Lieb-Robinson Hamiltonian (or that it obeys a Lieb-Robinson bound) if a bound of the form Eq.~(\ref{lrbound}) holds for some $v_{LR}$ and some exponentially decaying $g(l)$.
A parameter dependent family of Hamiltonians $H_s$ uniformly obeys a Lieb-Robinson bound if for some $v_{LR}$ and $g(l)$ the bound Eq.~(\ref{lrbound}) holds
for all $s$. Such a family is called ``uniformly LR''.
\end{definition}

We also want to define what it means for a Hamiltonian to have multiple ground states and a spectral gap.  Note that it is common practice in physics to refer to a system, such as a fractional quantum Hall system which has three low-lying states with an exponentially small splitting between them and then a gap to the rest of the spectrum as having a ``degenerate ground state", even though the non-vanishing splitting means that the lowest eigenvalue is in fact non-degenerate.   Our definition will reflect this usage, as we will not require that the states that we refer to as ``ground states" be degenerate.  All we will require is that the ``ground states" be separated from the rest of the spectrum by a gap.  

\begin{definition}
A Hamiltonian has $n$ ground states and a spectral gap $\Delta E$, if $E_{n-1}+\Delta E \leq E_{n}$ where the eigenvalues of the Hamiltonian
are $E_0,E_1,...$ with $E_0 \leq E_1 \leq ...$.
A family of Hamiltonians $H_s$ has $n$ ground states and a uniform spectral gap $\Delta E$ if $E_{n-1}(s)+\Delta E \leq E_n(s)$ for all $s$, where
the eigenvalues of $H_s$ are $E_0(s) \leq E_1(s) \leq ...$
\end{definition}

Note that we did not require in the above definition that the splitting $E_{n-1}-E_0$ between the different ground states be small.  For all the systems we are concerned with, this splitting will turn out to be small, but since it is not required to be small for lemma \ref{qadlemma}, we do not include this in our definition (in some applications of Lieb-Robinson bounds, the splitting between different ground states is important, but we don't need it here).

The next lemma expresses how the ground states evolve under changes in the Hamiltonian.  We consider some parameter dependent family
of Hamiltonians, $H_s$, for $0 \leq s \leq 1$, and imagine this family as describing some path from an initial Hamiltonian at  $s=0$ to
some final Hamiltonian at $s=1$.  
Stated roughly, this lemma shows that
if the Hamiltonian is gapped and local, then the change in the ground state under a local change in the Hamiltonian can be expressed by a local
operator acting on the ground state.  
\begin{lemma}
\label{qadlemma}
Let $H_s$ be a uniformly Lieb-Robinson family of Hermitian Hamiltonians, for $0\leq s \leq 1$, with $H_s$ differentiable with respect to $s$,
such that
$\partial_s H_s$ is supported in a disk $X$ of radius $R$ and such that for some $J$,
and $\Vert \partial_s H_s \Vert \leq J$ for all $s$. 
Let $H_s$ have uniform gap $\Delta E$.
Let $P(s)$ denote the Hermitian projector onto the ground state subspace of $H_s$.
Then, for any $l$, there exists a family of unitaries $U_s$ supported on the set of sites within distance $l$ of $X$ such that
\be
\label{Ubound}
\Vert U_s P(0) U_s^\dagger-P(s) \Vert
\leq {\rm const.}  \frac{J}{\Delta E} \Bigl(\exp(-l \Delta E/2v_{LR})+g(l)\Bigr).
\ee
\begin{proof}
The proof largely follows previous results on quasi-adiabatic continuation and is given in Appendix \ref{sec:appendix} for completeness.
\end{proof}
\end{lemma}

We make a few remarks.  First, note the appearance of  $g(l)$ in the lemma above.  For a Hamiltonian with exponentially decaying interactions, $g(l)$ will decay exponentially in $l$, but for a Hamiltonian with bounded range interactions, $g(l)$ will decay faster than exponentially in $l$, and the error will be dominated by  the term $\exp(-l \Delta E/2 v_{LR})$.  Further, in the case of exponentially decaying interactions,
the length scale over which $g(l)$ decays will be set by the decay scale of the interactions in the Hamiltonian, i.e., by the  microscopic details of the interaction rather than the magnitude of the spectral gap.

Note that in the lemma above, a bound on $\Vert \partial_s H_s \Vert$ appears; that is, the
bounds depend upon how rapidly the Hamiltonian changes along the path.  To give a physical explanation of why this appears,
consider dragging an anyon along some path.  Suppose we move the anyon a distance $L$.  Then, since we always scale the path length to unity
(that is, $s$ ranges from $0$ to $1$), the ``velocity" at which the anyon moves along the path is proportional to $L$.  Thus, for larger
$L$, we are moving the anyon more rapidly along the path, and so $\Vert \partial_s H_s \Vert$ will be larger; thus, in a sense the
appearance of $\Vert \partial_s H_s \Vert$ is really a way of measuring the distance we drag the anyons.  Thus, it is worth
restating the result in a re-scaled way: suppose that we drag an anyon a distance of order the disk radius $R$.  Then, typically we will have $\Vert \partial_s H_s \Vert \leq J$ for a $J$ of order $R$.  So, given that the error in Eq.~(\ref{Ubound}) is exponentially small in $l$, in such a case it suffices to choose $l$ logarithmically large in $R$ in order to make the error of order $1$.  For an $l$ of order $R$, the error will be exponentially small in $R$.

Also, note that if $\partial_s H_s$ is approximately supported in $X$, in that it can be approximated, up to exponentially small error,
by an operator supported in $X$, then we can derive a similar bound to Eq.~(\ref{Ubound}) which will involve the error in approximating $\partial_s H_s$.  We omit this case.

Finally, in the case that the Hamiltonian $H_s$ is a sum of commuting terms with bounded-range for all $s$, the Lieb-Robinson velocity is zero.  In this case, it is possible to show that for sufficiently large $l$, the error $\Vert U_s P(0) U_s^\dagger-P(s) \Vert$ is exactly zero.

\subsection{Proof of Absence}

Examined in detail, the ground state manifold $G^{\mathcal{G}}$ and the projector $P^{\mathcal{G}}$ which defines it depends on: 1) the number and location ($\Gamma$) within the $2$-sphere $S^2$ of the anyons, 2) the anyon particle type---a kind of boundary condition, and 3) the (possibly non-unitary) trivalent vertex normalization $f:  L^3\rightarrow \mathbb{C}\backslash 0$ or gauge choice, $L$ being the label set.  For Fib, DFib, and their Galois conjugates and time reversals (represented by $\bar{}\;$), $f$ is always symmetric and satisfies a consistency relation with the $F$-symbols: suppose $\{\til{F}^{ijk}_{l;nm} \}$ are new $6j$ symbols from $\{ F^{ijk}_{l;nm} \}$ by a gauge change $\{f(a,b,c)\}, a,b,c\in L$, then
$$ \til{F}^{ijk}_{l;nm}=F^{ijk}_{l;nm} \cdot \frac{f(j,k,n)f(i,n,l)}{f(i,j,m)f(m,k,l)}.$$

Except that it would unpleasantly cluster the notation, we should write $G^{\mathcal{G}}_{n,\Gamma, f}$ and $P^{\mathcal{G}}_{n,\Gamma, f}$.  The detailed position $\Gamma$ of the anyons within the lattice model is important to us since our proof will work with the entire "braid groupoid" $\overline{B_n}$.  
In fact, we treat $\Gamma$ as a continuous variable on a compact space of $2n$ (real) dimensions.  This moduli space of
anyon position is compact since distinct anyons are not permitted to closely approach.
The elements of $\overline{B_n}$ are oriented paths of {$n$-distinct (marked and framed) points in $\mathbb{R}^2$} which compose only when end points match.  $\overline{B_n}$ represents in a large but finite dimensional Hilbert space $H$ of microscopic degrees of freedom on $S^2$, the north pole serving as a standard $\infty$ for $\mathbb{R}^2$.  The vertex normalization $f$ is also important within the proof.  As we have already seen, the symmetric normalization yields a TQFT with all definite Hilbert spaces (though some are positive-definite and others are negative-definite).  The proof of Theorem \ref{theorema}  requires as a ''kernel" a single Hilbert space on which a non-singular form of mixed signs is preserved by $\overline{B_n}$.  With this kernel in hand, the proof actually covers all vertex normalizations $f$.

\begin{definition} We call an operator $\mL$ range $r$ if it is supported on a ball of diameter $r$.  Also, we use the same term
for sums of such operators. Similarly, an operator is called weakly range $r$ (in either sense) if it is range $r$ up to
exponentially small corrections.   We say that an operator is short range if it is supported on a ball of diameter small compared to system size.  

We say an operator $O$ is a local normalizer iff there is some constant $c$ which is small compared to system size such that $O\mL O^{-1}$ is range $r+c$ whenever $\mL$ is range $r$.  We say that an operator is a weakly local normalizer iff there is some constant $c$ which is small compared to system size such that $O \mL O^{-1}$ is weakly range $r+c$ whenever $\mL$ is range $r$.

A uniform family of (weakly) local normalizers $O_\Lambda$ is a parameter dependent family of operators such that $O_\Lambda \mL O^{-1}_\Lambda$ is (weakly) range $r+c$ whenever $\mL$ is (weakly) range $r$, with a uniform bound on the exponentially small corrections and on the constant $c$, and such that whenever $|\Lambda-\Lambda'| \leq {\cal O}(1)$, the product $O_\Lambda O^{-1}_{\Lambda'}$ is a product of at most ${\cal O}(1)$ operators which are all (weakly) range $r$ and are not necessarily the same, for some $r$ which is ${\cal O}(1)$.
An example of a local normalizer is a finite depth quantum circuit of invertible (not necessarily unitary) local operators.
An example of a uniform family of local normalizers is a family of finite depth quantum circuits of invertible local operators, such that an ${\cal O}(1)$ change in the parameter changes only ${\cal O}(1)$ different operators in the circuit; for the applications we have in mind, one should imagine that the parameter $\Lambda$ refers to different anyon positions and that changing $\Lambda$ changes the  circuit only near the anyon positions.

In the definition of weakly local normalizer, it will be important to define how we quantify the error term in the approximation by a bounded range operator.  The natural way to do this would be to require that the error term be small in operator norm compared to the operator norm of $O \mL O^{-1}$.  However, for technical reasons, for use later we will be interested in a what we call a g.s.-weakly local normalizer (g.s. stands for ground state).  In this case, we consider certain operators $M(i)$ which have the property that $M(i)$ is bounded range and exactly maps the ground state subspace of some non-Hermitian Hamiltonian to the ground state subspace of some other non-Hermitian Hamiltonian, with $M(i)^\dagger M(i)$ exactly preserving the ground state subspace of the first non-Hermitian Hamiltonian and having its ground state expectation value equal to its norm.  Then, we require that the error term be small in operator norm compared to the norm $|O_{\Lambda(i+1)} M(i) O_{\Lambda(i)}^{-1}\psi|$ for $\psi$ in the ground state of some other Hermitian Hamiltonian (this ground state subspace is obtained by applying $O$ to the ground state subspace of the Hermitian Hamiltonian).  Note that if $O$ were an isometry, then the norm $|O_{\Lambda(i+1)} M(i) O_{\Lambda(i)}^{-1} \psi|$ would equal the norm of $M$ and so this would reduce to the more natural definition.  Note also that any local normalizer is a g.s.-weakly local normalizer.
\end{definition}

\begin{theorem}
\label{theorema}
Fixing the number $n\geq 5$ and particle type $\tau\otimes \tau$ of DFib anyons on $S^2$ and any vertex normalization $f$ there can be no continuous uniform $\Gamma$-family of (g.s.-weakly) local normalizer operators $O_{\Gamma}$: ${\cal H}\rightarrow {\cal H}$, so that $\mathcal{O}_\Gamma G^{\mathcal{G}}_{n,\Gamma, f}$ is, for all anyon positions $\Gamma$ the ground state manifold of a uniformly Lieb-Robinson and uniformly gapped family of Hermitian Hamiltonians $H(\Gamma)$ defining a topological phase [see Eq. (\ref{eq:code})].
\end{theorem}

\textit{Proof.}  The theorem uses the notation of reference 21 to describe the anyons in DFib.  For now, fix the algebraic vertex normalization $\lambda=f$.  Below we may suppress $\Gamma$ and $f$ from the notation when they play no role.

Suppose $\mathcal{O}_\Gamma$  exists, then$\mathcal{O}_\Gamma G^{\mathcal{G}}_{\Gamma}$ is a family of code subspaces and for $\Gamma$ near $\Gamma'$ the subspaces are connected up to exponentially small discrepancy by a local unitary $U_{\Gamma,\Gamma'}$ (these are the $U_s$ of Lemma \ref{qadlemma}.
Writing $\textrm{DFib}_f^{\mathcal{G}}\cong \textrm{Fib}_f^{\mathcal{G}}\otimes {\overline{\textrm{Fib}}_f^{\mathcal{G}}}$ (one may think DFib describes a bilayer), let us recall a theorem stated in \onlinecite{GK} for the right hand factor $\textrm{Fib}_f^{\mathcal{G}}$, where $f$ is the algebraic normalization (Note: while $\textrm{DFib}_f^{\mathcal{G}}$ is a theory of string-nets on the surface $S^2$, with boundary conditions at anyons, $\textrm{Fib}_f^{\mathcal{G}}$ is the corresponding string-net theory\cite{Freedman00} in the $3$-ball with boundary $S^2$.   Thus the function $f$ gauging vertices acts compatibly in both theories.)

%%%INSERT 2
Now, according to Ref.~\onlinecite{GK}, corollaries 1.2.4 and 1.2.6, for n $\geq$ {5}, the Jones representation $\rho$ on the topologically defined Hilbert space $V_R$ of ground states for $\rm{Fib^{\mathcal{G}}}$ is (analytically) dense in a noncompact special unitary group (preserving a Hermitian metric of mixed signs) $SU(p,q)$ := $SU((X,n\cdot{1},0,e^{\frac{4\pi{i}}{5}}))\cong{SU((X,n\cdot{2},0,e^{4\pi{i}/5}))}$.

Recall that in subsection \ref{Sec:GaloisConjugate} we confirmed that the loop value for a closed string in $\textrm{DFib}_f^{\mathcal{G}}$  was $-\frac{1}{\phi}$ and that with the algebraic vertex normalization the signs of the $\overline B_n$-invariant ground state "Hilbert" spaces are indeed mixed: $p>0$ and $q>0$.

We need to formulate a lemma regarding the following concept.

\begin{definition}  An ambient groupoid representation is a functor from a groupoid to the category of subspaces of a fixed Hilbert space, and linear transformations on $H$ carrying one image subspace to another.
\end{definition} 

Thus to objects $a,b$ of the groupoid we assign spaces $A \subset H, B\subset H$ and to a morphism $a\rightarrow b$ a unitary map $H\rightarrow H$ carrying $A$ to $B$. In our case, the groupoid is $\overline{B}_n$ with $\Gamma$ being the objects and motions of anyons being the morphisms.  The subspaces are the respective ground states for $H_{LW,alg.}^{\mathcal{G}}$---the Galois conjugated Levin-Wen Hamiltonian with algebraic vertex normalization.

\begin{lemma}
\label{lemma}
Let $\{A_i\}, i\in \Gamma$ be the set of code subspaces of fixed Hilbert space $H$ and $\overline{B}=\{\Gamma, \{\rightarrow\}\}$ a groupoid.  Suppose for a generating set of morphisms in $\overline{B}$, $a \stackrel{\theta}{\rightarrow} b$, there are invertible local operators $\mathcal{L}_{\theta}: H\rightarrow H$ with $\mathcal{L}A=B$ ($A$=image $a$ and $B$=image $b$).  Then this data determines a unique projective ambient groupoid representation .  That is, all compositions commute with the functor up to multiplication by a nonzero scalar.

\begin{proof}
 Consider $A\stackrel{\mathcal{L}_{ \theta}|_A }{\longrightarrow} B \stackrel{\mathcal{L}_{\phi}|_B}{\longrightarrow} C$, $\mathcal{L}_{\theta}$ and $\mathcal{L}_{\phi}: H\rightarrow H$ being the local operators carrying $A$ to $B$ and $B$ to $C$ respectively.  The composition $\mathcal{L}_{\phi}\cdot \mathcal{L}_{\theta}$ is local and $\mathcal{L}_{\phi}\cdot \mathcal{L}_{\theta} (A)=C$.  By the code property there is a scalar $\lambda$ so that $\Pi_A (\mathcal{L}_{\phi}\cdot \mathcal{L}_{\theta})^{-1} \cdot (\mathcal{L}_{\phi}\cdot \mathcal{L}_{\theta}) \cdot \textrm{inc}_A=\lambda \cdot \textrm{Id}$, where $\lambda\neq 0$ since all morphisms $\mathcal{L}$ in the representation are invertible.  ($\Pi_A$ denotes the appropriate projection to $A$ corresponding to its code property.) Thus (projectively) the homomorphic property, restricted to the subspaces $\{A_i\}, i\in \Gamma$, is redundant when these subspaces have the code property.

For uniqueness consider two possible representations $\mathcal{L}(\theta)$ and $\mathcal{L}'(\theta)$.  $(\mathcal{L}')^{-1}\mathcal{L}: H\rightarrow H$ is local and carrying $A$ to itself.  Thus $\Pi_A(\mathcal{L}'_{\theta})^{-1} \mathcal{L}_{\theta}\cdot \textrm{inc}_A: A\rightarrow A$ is multiplication by some scalar $\lambda$ which, again by the invertibility of $\mathcal{L}_{\theta}$ and $\mathcal{L}'_{\theta}$, is non-zero.
\end{proof}
\end{lemma}

By assumption the collection $\{\mathcal{O}(G^{\mathcal{G}}_{\Gamma}) \}$ is a code with respect to the usual Hermitian projection. Using the $U_s$ of Lemma \ref{qadlemma} as the generating set of morphisms, Lemma \ref{lemma} builds a projective representation $\brho: \overline{B_n}\times H \rightarrow H$ of DFib up to exponentially small errors which we can neglect for the moment but return to shortly.  We may think of this
representation as the result of (quasi-)adiabatic evolution of $G_\Gamma^{\mathcal G}$ inside the finite dimensional microscopic Hilbert space $H$.  Intuitively, braiding might be realized by building and slowly moving a potential trap term added to the Hamiltonian.  For plaquette excitations (e.g. $\tau\otimes \tau$) such a trap could have the rough form $H_{\textrm{trap}}=arctan(\delta t)B_p + (\pi - arctan(\delta t) B_{p'}+\epsilon \sigma_i^z$ for plaquettes $p$ and $p'$ separated
by edge.  Such family of Hamiltonians will adiabatically braid the anyons.  Formally, however, we have posited, for contradiction,
the family $H(\Gamma)$ and this is all we need.

Lemma \ref{qadlemma} provides  local unitaries intertwining $G^{\mathcal{G}}_{\Gamma_1}$ and $G^{\mathcal{G}}_{\Gamma_2}$, and so  $\brho$ is obviously a
(projective) unitary representation with respect to the standard positive definite Hermitian structure on the space $H$ of microscopic degrees of freedom.  Thus $\brho$ preserves ordinary lengths and angles (as measured in $H$) and the complex structure (multiplication by $i$) of $H$ as well.
So this $\brho$ manufactured from Lemma \ref{lemma} looks geometrically quite distinct from a second representation of $\overline{B_n}$, $\brho':=\mO (\rho\otimes \rho^{*}) \mO^{-1}$. As explained above (also see ref. \cite{Freedman00}), $\rho$ and $\rho*$ also act on the same collection of string-net spaces as $\brho$ and the uniqueness clause of  Lemma \ref{lemma} yields a projective isomorphisms $\brho \stackrel{proj}{\cong} \brho'$.

We learned from Ref. \onlinecite{GK} that $\rho$ is dense in $SU(p,q), p>0, q>0$, and that $X(n\cdot 2, 0, e^{\frac{4\pi i}{5}})$ is the fundamental representation $\omega_1$ of $SU(p,q)$.
%%%INSERT 4
$\omega_1$ preserves a form of {\em mixed} signs and thus will distort not only Euclidean length but Euclidean angles as well by an unbounded amount (For example consider the effect of boosts in $O(1,1)\subset U(1,1)$ on Euclidean angle). This difference, that $\brho$ and $\brho'$ preserve forms of different signatures, excludes the existence of a (g.s.-weakly) local normalizer $\mO$ transforming $\{G^{\mathcal G}_{\Lambda}\}$ to the ground state spaces of $H(\Lambda)$.

We now
consider in more detail the exponentially small errors that we have neglected.   First some preliminaries.  We will need the following lemma which provides a useful corollary of the disk axiom:
\begin{lemma}
\label{diskcorollary}
Let $X$ be some set and $P$ some projector such that for any operator $A$ supported on $X$ there is a scalar $z$ such that
\be
\label{axiom}
\Vert P A P - z P \Vert \leq \epsilon \Vert A \Vert,
\ee
for some sufficiently small $\epsilon$.  (In typical applications, we have in mind that $P$ is the projector onto the ground state subspace of some system, $X$ is some set of small diameter, and the above equation encodes a soft form of the disk axiom for that theory).
Then,
for any operator $O$ supported on set $X$ there is a scalar $w$ such that
\be
\Vert POP-w P \Vert \leq C \sqrt{\epsilon \Vert (1-P) O P \Vert \, \Vert P O (1-P) \Vert},
\ee
for some constant $C$ of order unity (the constant $C$ is independent of $\epsilon$, so long as $\epsilon$ is sufficiently small).  
\begin{proof}
First, assume that $O$ is Hermitian.  Consider the operator $U=\exp(it O)$, for $t$ real.  Since $\Vert U \Vert=1$,
$\Vert PUP-zP \Vert \leq \epsilon$ for some $z$.  We can expand $PUP$ in a power series in $(1-P)OP$  giving
\be
PUP=P \exp(i t POP) P +{\cal O}(t^2 \Vert P O (1-P) \Vert^2).
\ee
Suppose $POP=w+\Delta$, for some traceless operator $\Delta$ and scalar $w$.  Pick $t=2\epsilon/\Vert \Delta \Vert$.  Then,
\be
PUP=Pw+2i\epsilon P\frac{\Delta}{\Vert \Delta \Vert} P+{\cal O}(\epsilon^2+\epsilon^2\Vert PO(1-P) \Vert^2/\Vert \Delta \Vert^2).
\ee
So, the distance from $PUP$ to the closest scalar multiple of $P$ is at least $2\epsilon-{\cal O}(\epsilon^2+\epsilon^2 \Vert P O (1-P) \Vert^2 / \Vert \Delta \Vert^2)$.
If $\Vert \Delta \Vert$ is sufficiently large compared to $\sqrt{\epsilon} \Vert PO(1-P) \Vert$, then the term ${\cal O}(...)$ is small compared to the leading term, assuming $\epsilon$ is sufficiently small (we need $\epsilon$ sufficiently small compared to unity so that term ${\cal O}(\epsilon^2)$ is small).  However, this contradicts the assumption that $PUP$ is within distance $\epsilon$ of some scalar multiple of $P$.

Now, consider the general case that $O$ need not be Hermitian.  Add an additional spin-$1/2$ degree of freedom to the system and consider the Hermitian operator $\tilde O=(O-w)\otimes\sigma^++(O^\dagger-\overline w)\otimes\sigma^-$, where $\sigma^+,\sigma^-$ are the
raising and lowering operators for that spin and $w$ is chosen so that $P(O-w)P$ is traceless.  The assumption (\ref{axiom}) for the original theory implies that for the system with the additional spin-$1/2$, for any operator $A$ acting on set $X$ and on the additional spin-$1/2$ that
\be
\Vert PAP-P\otimes Q \Vert \leq 4 \epsilon \Vert A \Vert,
\ee
for some $2$-by-$2$ matrix $Q$ acting on the additional spin (to show the above equation, expand $PAP$ as a sum of four product operators, one operator in the product acting on $X$ and the other on the added spin, and apply Eq.~(\ref{axiom}) to each term in the product).  Construct $U=\exp(i t \tilde O)$.  Perturbatively expand the $\sigma^+$ component of $PUP$ for the same $t$ as before.  This is
\be
2i\epsilon P \frac{\Delta}{\Vert \Delta \Vert}P\otimes \sigma^++{\cal O}(\epsilon^3+\epsilon^2\Vert PO(1-P)\Vert \Vert (1-P)OP\Vert/\Vert \Delta\Vert^2).
\ee
We again get a contradiction as in the Hermitian case.
\end{proof}
\end{lemma}

The representation $\brho$ gives a mapping from braids to matrices.  We will use $M$ to refer to such a matrix.  We can also construct a matrix $U$ by taking the $U_s$ of lemma \ref{qadlemma} for the corresponding braid and projecting into the grounds state subspace.  So long as the length of the braid is smaller than some quantity growing exponentially with the linear size of the system, the matrix $U$ will be approximately unitary (the error arises from leakage out of the ground state subspace in lemma \ref{qadlemma}.  We claim that
\begin{lemma} 
\be
\Vert U-z \mO M \mO^{-1} \Vert<<1
\ee
 for such braids for some scalar $z$.  Indeed, the difference in norms is exponentially small in system size.
\begin{proof}
This difference follows from the disk axiom: we decompose the braid into $n$ segments; taking $n$ of order $L$, each segment moves the anyons only a short distance. In the non-unitary theory, we can construct short range operators which move the ground state subspace before the $i$-segment to the ground state subspace after that segment. Call these operators $M(i)$, so that $M$ is equal to the projection of $M(n)...M(1)$ into the ground state subspace.  This operator $M(n)...M(1)$ exactly preserves the ground state subspace of the Galois conjugated theory and $\mO M \mO^{-1}$ exactly preserves the ground state subspace of the
unitary theory.  We use $P(i)$ to denote the projector onto the ground state subspace after the $i$-th segment, and we use \be
M'(i)=\mO_{\Lambda(i)} M(i) \mO_{\Lambda(i-1)}^{-1},
\ee
where $\mO_{\Lambda(i)}$ is the operator $\mO_\Lambda$ corresponding to the position of the anyons after the $i$-th segment.
so that $M(i)P(i-1)=P(i) M(i)$ and $P(n)=P(0)$.
In the unitary theory, let $U(i)$ denote the unitary matrices from lemma \ref{qadlemma} for the motion along the $i$-th segment, so that $U$ is equal to the projection into the ground state subspace of $U(n)...U(2)U(1)$.  Note that $U(n)...U(1)$ preserves the ground state subspace up to exponentially small exponentially small error and the matrices $U(i)$ are bounded range.  Thus,
\begin{eqnarray}
\label{approxprod} && \Vert U-z \mO M \mO^{-1} \Vert \\ \nonumber
&=&\Vert P(0)U(n)U(n-1)...U(1) P(0) \\ \nonumber
&&-zP(0) M'(n)M'(n-1)...M'(1)P(0)\Vert \\ \nonumber
&\approx & \Vert P(0) U(n) P(n-1) U(n-1)P(n-2) ... U(1) P(0) \\ \nonumber
&&-z P(0) M'(n)  P(n-1) M'(n-1) P(n-2)...   M'(1) P(0) \Vert,
\end{eqnarray}
where the left side of this approximate equality differs from the right by inserting additional factors of $P(i-1)$ after every occurrence of
$U(i)$ or $M'(i)$ on the right-hand side.   We have $P(0) M'(n) M'(n-1) ... M'(1) P(0)=P(0) M'(n) P(n-1) M'(n-1) P(n-2) ... M'(1) P(0)$, but
the error in Eq.~(\ref{approxprod}) occurs because $U(i) P(i-1)$ is only approximately equal to $P(i) U(i-1)$ (the difference is exponentially small in
system size).  Note that on the right-hand side of this equation the matrices act in the ground state Hilbert space, while on the left-hand side they act in the full Hilbert space.
To bound the right-hand side of Eq.~(\ref{approxprod}), it suffices to show that, for all $i$,
\be
\label{estneeded}
\Vert P(i) U(i) P(i-1) - z(i) P(i) M'(i) P(i-1) \Vert
\ee is exponentially small for some scalar $z(i)$.  To see this, set
\be
z=\prod_i z(i)
\ee
So,
\begin{eqnarray}
&& \Vert P(0) U(n) P(n-1) U(n-1) ... P(1) U(1) P(0)-
\\ \nonumber &&P(0) M'(n) P(n-1) M'(n-1) ... P(1) M'(1) P(0) \Vert \\ \nonumber
&\leq &
\Vert P(0) U(n) P(n-1) ... P(1) U(1) P(0)- \\ \nonumber
&& P(0) U(n) P(n-1) z(n-1) M'(n-1)... z(1) M'(1) P(0) \Vert \\ \nonumber
&&+\Vert P(n) U(n) P(n-1)-P(n) z(n) M'(n) P(n-1) \Vert \\ \nonumber
&&\times \prod_{i=1}^{n-1} \Vert P(i) M'(i) P(i-1) \Vert
\\ \nonumber
&\leq& ...
\end{eqnarray}
Given this norm estimate (\ref{estneeded}), then since  $P(i) U(i) P(i-1)$ is an  approximate isometry from the range of $P(i-1)$ to the range of $P(i)$, the matrix
$P(i) z(i) M'(i) P(i-1)$ is also such an approximate isometry, so the product of norms $\Vert P(i) M'(i) P(i-1) \Vert$ above is bounded.

So, we must bound Eq.~(\ref{estneeded}).  Since $Ui)$ is an approximate isometry, it suffices to bound
$\Vert P(i-1)-P(i-1) U(i)^\dagger z(i) M'(i) P(i-1) \Vert$.  At first sight, this seems to follow immediately from the disk axiom: since $U(i)^\dagger M'(i)$ is short range, or at least approximately short range (note that for $M'(i)$ this follows by the definition of a family of local normalizers, but see the next paragraph for a more careful treatment of error terms), by the disk axiom it is close to a scalar when projected into the ground state subspace.  Hence, choosing $z(i)$ to be the inverse of this scalar, the desired result seems to follow.  However, there is a complication: suppose $P(i-1) U(i)^\dagger M'(i) P(i-1)$ is within some distance $\epsilon$ of  $z(i)^{-1} P(i-1)$ for some $z(i)$; then, we bound
$\Vert P(i-1)-P(i-1) U(i)^\dagger z(i) M'(i) P(i-1)\leq \epsilon  |z(i)|$.  Hence, if $z(i)$ is large, the resulting error can be large even if $\epsilon$ is small.  This is why we will need the lemma (\ref{diskcorollary}) above.

By definition of g.s.-weakly local normalizer, the operators $M'(i)$ can be approximated by operators that are short range, up to an error that is small compared to $| M'(i) \psi|$ for all $\psi$ in the ground state subspace $P(i-1)$ with $|\psi|=1$.  Since $U(i)$ is approximately unitary and an approximate isometry between two ground state subspaces, this means that $U(i)^\dagger M'(i)$ can be approximated by an operator $O(i)$ that is short range, up to an error that is small compared to $| U(i)^\dagger M'(i) \psi|$ for $\psi$ in $P(i-1)$.  (Note that if $\mO$ is a local normalizer, then $M'(i)$ already is short range so we can take $O(i)=U(i)^\dagger M'(i)$ in that case.)  So, for the $O(i)$,  $\Vert (1-P(i-1)) O(i) P(i-1) \Vert$ is small compared to $|O(i) \psi|$ for all $\psi$.  Applying lemma (\ref{diskcorollary}), this means that
$P(i-1) O(i) P(i-1)$ is close to $z(i) P(i-1)$, for some $z(i)$ up to an error that is small compared to $z(i)$.
\end{proof}
\end{lemma}

We can find a braid such that the corresponding matrix $M$ is diagonalizable, and with the ratio between its largest and smallest eigenvalue being at least $2$ in absolute value\cite{nonunitarySK}.
This means that the ratio between the largest and smallest eigenvalue of $z \mO M \mO^{-1}$ is at least $2$ in absolute value.  However, $z \mO M \mO^{-1}$ is close to a unitary matrix.  All eigenvalues of a unitary matrix are on the unit circle in the complex plane.  Further, a small perturbation of a unitary matrix leaves all of its eigenvalues close to the unit circle, so $z \mO M \mO^{-1}$ must have all of its eigenvalues close to the unit circle contradicting the assumption on the ratio of eigenvalues.

Now we remove the condition on the vertex normalization or "choice of gauge".  As noted as above the gauge choice is a function $f: L^3\rightarrow \C \backslash 0$.  Although this is not crucial, it is pleasant that in the Fib case $f$ is identically $1$ except for taking value $f(\tau, \tau, \tau)=\lambda$ on the essential trivalent vertex.  Recalling the $A_v$ and $B_p$ terms in the Levin-Wen model Hamiltonian, the fusion rule terms, $A_v$, commute with and do not depend on $f$.  The effect of $f$ on $B_p$ may be computed (using the compatibility of gauge choice and the $F$-matrices used to construct $B_p$):
\begin{equation}
\label{eq:star}
B_{p,\lambda}=F_{\lambda}\cdot B_{p,alg.}\cdot F_{\lambda}^{-1},
\end{equation}
where $F_{\lambda}$ is the ``relative fugacity matrix", $F_{\lambda}: H\rightarrow H$.  $F_{\lambda}$ is diagonal in the string-net basis $\{k\}$ and the entry $F_{k,k}$ is simply $\lambda^{\# (k)}$, where $\#(k)$ denotes the number of $(\tau, \tau, \tau)$-vertices in the $k^{th}$ string-net.

As a concrete example of the above gauge dependence of $B_p$ in the Fib case
\begin{equation}
F^{\tau \tau \tau}_{\tau; sym}=\begin{pmatrix}
1&0\\0& \lambda^{-2} 
\end{pmatrix}
F^{\tau \tau \tau}_{\tau; alg}
\begin{pmatrix}
1&0\\0& \lambda^2
\end{pmatrix}.
\end{equation}
By inspecting the two $F$-matrices in the two normalizations, we have $\lambda^2=i\phi^{5/2}$.
Given a trivalent graph $\gamma$, and a labeling of its edges $k_{\gamma}$.  Let $\{\Psi_{alg., k_{\gamma}}\}$ and $\{\Psi_{sym., k_{\gamma}}\}$ be the two basis of the Hilbert space $\otimes_{e\in \gamma} \C^{\#(L)}$ of labeled graphs.  Then $\Psi_{sym., k_{\gamma}}=\lambda^{\#(k_{\gamma}})\Psi_{alg., k_{\gamma}}$.  Suppose 
\begin{equation}
B_{p,alg.}\Psi_{alg., k_{\gamma}}=\sum_{k'_{\gamma}} B_{p,alg.,k'_{\gamma},k_{\gamma}} \Psi_{alg., k'_{\gamma}},
\end{equation}
 noting that gauge change and recoupling are commutative, we obtain
\begin{equation}
B_{p,sym.}\lambda^{-\#(k_{\gamma})} \Psi_{sym., k_{\gamma}}=\sum_{k'_{\gamma}} B_{p,alg.,k'_{\gamma},k_{\gamma}} \lambda^{-\#(k'_{\gamma})}\Psi_{sym., k'_{\gamma}}.
\end{equation}
Thus $B_{\Gamma,sym}$ has the claimed conjugated form. Similarly for any vertex fugacity $\lambda$, Eq. (\ref{eq:star}) holds. Thus for a general relative fugacity $\lambda$, $\mathcal{G}_{\Gamma,\lambda}=F_\lambda \mathcal{G}_{\Gamma,alg.}$

Observe that while $F_{\lambda}$ is not local (in the sense of having supported on a disk of bounded radius), both $F_\lambda$
and $F_\lambda^{-1}$ are implemented by a depth $1$ invertible circuit, $F_\lambda=\prod_{\rm sites}(\lambda \Pi_{\tau\tau\tau}+(1-\Pi_{\tau\tau\tau}))$.  Thus the general local normalizer operator $\mO$ may be written as
$\mO=\mO' \circ F_\lambda$, where $\mO'$ is also local normalizer; given $\lambda, \mO$, and $\mO'$ determine each other uniquely, we have just shown that for all local normalizer $\mO$ there is a local $\mL$ so that $\mL$ acts on $\mO G^{\mathcal G}_{\rm alg}$, i.e. that
\be
{\rm inc}^\dagger_{G^{\mathcal G}_{\rm alg}} \circ \mO^\dagger \circ \mL \circ \mO \circ {\rm inc}_{G^{\mathcal G}_{\rm alg}} \neq {\rm scalar}.
\ee
It follows that for all $\mO'$ there is an $\mL$ (identical to $\mL$ above) so that:
${\rm inc}_{G^{\mathcal G}_{\rm alg}} \circ F^\dagger_\lambda \circ \mO'^\dagger \circ \mL \circ \mO' \circ F_\lambda \circ {\rm inc}_{G^{\mathcal G}_{\rm alg}} \neq {\rm scalar}$.
But
\be
{\rm inc}_{G^{\mathcal G}_{\rm alg}}\circ F_\lambda^\dagger \circ \mO'^\dagger \circ \mL \circ \mO' \circ F_\lambda \circ {\rm inc}_{G^{\mathcal G}_{\rm alg}} = {\rm inc}_{G^{\mathcal G}_{\lambda}} \circ \mO'^\dagger \circ \mL \circ \mO' \circ {\rm inc}_{G^{\mathcal G}_\lambda}.
\ee
So we find that for the change of variables $\mO'$, $\mL$ acts on $\mO'\circ {\rm inc}_{G^{\mathcal G}_\lambda}$.  Varying $\mO$ over
all local normalizer operators produces a local normalizer $\mO'=\mO \circ F_\lambda^{-1}$.  Thus for every possible local normalizer operator $\mO'$ there is a local $\mL:H\rightarrow H$ acting on $\mO' G^{\mathcal G}_\lambda$.

This completes the proof of Theorem \ref{theorema} by removing the hypothesis of algebraic vertex normalization.
 \qed \\ \\ 
 
Theorem \ref{theorema} immediately implies the following corollary
\begin{corollary}
\label{theoremb}
Let $\{G_{\Gamma,f}^{\mathcal{G}} \}$ be the ground state manifolds for the Galois conjugated Levin-Wen Hamiltonian $H^{\mathcal{G}}_{\Gamma,f}$ for $n\geq 5$ $\tau \otimes \tau$-anyons on the $2$-sphere $S^2$ with positions $\Gamma$ and any vertex normalization $f$, within a larger Hilbert space $H$ of microscopic (lattice) degrees of freedom.  There can be no continuous uniform family of local normalizer  operators $\mO_\Gamma$ so that $\{\mO_\Gamma H^{\mathcal{G}}_{\Gamma,f} \mO_\Gamma^{-1}\}$ are uniformly  gapped uniformly Lieb-Robinson Hamiltonians determining topological ground states $\{\mO_\Gamma G_{\Gamma,f}^{\mathcal{G}} \}$, in the sense of TQO-1 \cite{BHM} ({\it i.e.} satisfying the code property (\ref{eq:code}) ).
\end{corollary}

Although we have concentrated the discussion on the Fib TQFT, its quantum double, and their Galois conjugates, the proof requires only two ingredients: 1) finding pairs of Galois conjugate theories (with choice of vertex gauge) one of which is unitary (for the Hilbert space of a sphere or plane with fixed anyon content) and one of which is unitary with respect to a Hermitian metric of mixed signs $(p,q), p>0, q>0$, and 2) establishing denseness of the braid group representations in $SU(p,q)$.  Using just the results for $SU(2)$-theories obtained in Ref. \onlinecite{GK}, infinitely many other unitary theories arise which have Galois conjugates satisfying Theorem \ref{theorema}.

\section{Comments on Non-Hermitian Hamiltonians}
While the bulk of our paper is devoted to showing that certain wave functions cannot be the ground states of gapped Hermitian Lieb-Robinson Hamiltonians, it is worth briefly discussing what is known about non-Hermitian Hamiltonians.  For non-Hermitian Hamiltonians, many of the technical tools involving Lieb-Robinson bounds are unavailable, and so many results that we know in the Hermitian case are not known here.

The first major difficulty in the non-Hermitian case is that even if the Hamiltonian is a sum of terms $H_Z$ which obey Eq.~(\ref{Hdecay}), if the Hamiltonian is not Hermitian, then we do not know if the Lieb-Robinson bound holds.  Similarly, for a Hermitian Hamiltonian, the Lieb-Robinson bound might not hold for evolution in imaginary time.
Given that the Lieb-Robinson bound fails, we are also unable to prove locality of correlation functions in a non-Hermitian Hamiltonian even if there is a gap in the spectrum.

The fact that we cannot prove locality of correlation functions is relevant to the following application of the disk axiom.  Suppose we have a Hermitian Hamiltonian which obeys the disk axiom with $P$ being the projector into the ground state subspace.  Suppose operators $O_X$ and $O_Y$ are supported on small disks $X$ and $Y$ such that the disk axiom implies that $P O_X P$ and $P O_Y P$ are both close to scalar multiples of $P$.  Now, let us ask whether the operator $P O_X O_Y P$ is also close to a scalar multiple of $P$.  Consider the case in which the disks $X$ and $Y$ are far separated such that the smallest disk containing both $X$ and $Y$ is too large to directly apply the disk axiom to $O_X O_Y$.  Thus, the disk axiom alone does not tell us that $O_X O_Y$ is close to a scalar when projected into the ground state subspace.  However, if we have a gapped, Hermitian, Lieb-Robinson Hamiltonian then correlations decay exponentially in any ground state, so that $P O_X O_Y P$ is close to $P O_X P O_Y P$, and then applying the disk axiom to $P O_X P$ and $P O_Y P$ implies that 
$P O_X O_Y P$ is close to a scalar multiple of $P$.  Unfortunately, though, in the non-Hermitian case we do not know that there is exponential decay of correlation functions, and so even if we assume the disk axiom for small disks, we do not see how to prove that $O_X O_Y$ is also close to a scalar when projected into the ground state subspace of a non-Hermitian Hamiltonian obeying the disk axiom.
In fact, suppose we consider two states $\psi_1,\psi_2$ such that any operator supported on a small disk is equal to a scalar when projected into the space spanned by $\psi_1,\psi_2$.  Let us relax any requirement that $\psi_1,\psi_2$ be ground states of a Hamiltonian, whether Hermitian or not, and simply take them to be arbitrary states.  Then, we can give an example in which product operators of the form $O_X O_Y$ given above are {\it not} close to a scalar when projected into the ground state subspace, even though both $O_X,O_Y$ are close to scalars when projected into this subspace, by considering a quantum error-correcting code on a small number of qubits, and defining the two states $\psi_1,\psi_2$ on the large lattice by placing the qubits defining the code on far separated sites of the lattice, and placing all other qubits in a product state.

\section{Conclusions}

To summarize, we have shown that a large class of non-unitary topological quantum field theories cannot be realized as ground states of Hermitian  (quantum mechanical) Hamiltonians that satisfy a Lieb-Robinson bound. This includes but is not limited to local and quasi-local (exponentially decaying) Hamiltonians. While our proof has been formulated for quantum doubles of TQFTs, it also rules out the realization of the constituent non-doubled TQFT in a Hermitian system: if the latter were to exist, it could be used to trivially construct a Hermitian model for the corresponding quantum double. 

The TQFTs covered by our proof include, in particular, the Galois conjugates of Fib and su(2)$_k$ TQFTs for $k=3$ and all $k\ge 5$. Among these, one case of special recent interest is the Yang-Lee TQFT (the Galois conjugate of Fib) underlying the proposed Gaffnian quantum Hall wave function. Our argument implies that this Gaffnian wave function cannot occur as ground state of a gapped fractional quantum Hall state (described by a Hermitian Hamiltonian), if one considers that the screened Coulomb interaction satisfies a Lieb-Robinson bound.

\section*{Acknowledgments}

We acknowledge discussions with E. Ardonne, B. Bauer, A. Ludwig, and K. Walker. We thank the Aspen Center for Physics. Our simulations used some of the ALPS libraries \cite{ALPS2,ALPS13} and partly also the ARPACK library. \cite{ARPACK} Data evaluation has been performed using the ALPS libraries and the VisTrails scientific workflow and provenance management system.\cite{VisTrails} Full provenance information and workflows to recreate the figures are available by following the hyperlinks associated with each figure.

\appendix

\section{Proof of Lemma \ref{qadlemma}}

 \label{sec:appendix}
We define the ``quasi-adiabatic continuation operator" ${\cal D}_s$ by
\be
i {\cal D}_s=\int {\rm d}t F(t) \exp(i H_s t) \Bigl(\partial_s H_s\Bigr) \exp(-i H_s t),
\ee
where the function $F(t)$ is defined by
(we follow \cite{hall}, while a more complicated choice of $F(t)$ was used in \cite{lsm,hwen}):
\be
\label{Fdef}
\frac{i}{\alpha\sqrt{2\pi}}\int_t^{\infty} {\rm d}u \exp(-u^2/2\alpha^2),
\ee
for $t>0$ and $F(t)$ for $t<0$ is defined by $F(t)=-F(-t)$.  
The quantity $\alpha$ in Eq.~(\ref{Fdef}) is some constant chosen below.

We now show that $\partial_s P(s)$ is close to $i[{\cal D}_s,P(s)]$, and bound the difference between the two expressions in operator norm.
Define $\tilde F(\omega)$ to be the Fourier transform of $F(t)$.
One may show that
\be
\label{omegaerr}
|\omega| \geq \Delta E \rightarrow |\tilde F(\omega)-1/\omega)| \leq {\rm const.} \times (1/\Delta E) \exp(-\alpha^2 \omega^2/2).
\ee
Let $\psi_i(s)$ denote eigenvectors of $H_s$ with eigenvalues $E_i(s)$, so $P(s)=\sum_{i=0}^{n-1} |\psi_i(s)\rangle \langle \psi_i(s)|$.
Then,
\begin{widetext}

\begin{eqnarray}
\label{psds}
i[{\cal D}_s,P(s)]&=&
\sum_{i=0}^{n-1} \;\; \sum_{j\geq n} \; \;
\int {\rm d}t F(t) |\Psi^j(s)\rangle \Bigl(\langle \Psi^j(s)| \exp(i H_s t)\partial_s H_s \exp(-i H_s t)|\Psi^i(s)\rangle\Bigr) \langle \Psi^i(s)|+h.c. \\ \nonumber
&=&
\sum_{i=0}^{n-1} \;\; \sum_{j\geq n} \; \;
\tilde F(E_i-E_j) |\Psi^j(s)\rangle \Bigl(\langle \Psi^j(s)| \partial_s H_s|\Psi^i(s)\rangle\Bigr) \langle \Psi^i(s)|+h.c.
\end{eqnarray}
By linear perturbation theory,
\begin{eqnarray}
\label{diff}
\partial_s P(s)&=&
\sum_{i=0}^{n-1} \;\; \sum_{j\geq n} \; \;
\frac{1}{E_i-E_j} |\Psi^j(s)\rangle \Bigl(\langle \Psi^j(s)| \partial_s H_s |\Psi^i(s)\rangle\Bigr) \langle \Psi^i(s)|+h.c.
\end{eqnarray}
Thus, by Eqs.~(\ref{omegaerr},\ref{psds},\ref{diff})
\be
\label{eq1}
\Vert \partial_s P(s)-[i {\cal D}_s,P(s)] \Vert \leq {\rm const.} \times (\Vert \partial_s H_s \Vert/\Delta E) \exp(-\alpha^2\Delta E^2/2).
\ee
\end{widetext}

We also have a bound on the time decay of $F(t)$:
\be
\label{Fdecay}
|F(t)| \leq {\rm const.}\times \exp(-t^2/2\alpha^2).
\ee
We now define $i{\cal D}_s^l$ to be an approximation to $i{\cal D}_s$ supported on the set of sites within distance $l$ of $X$.  To construct
this approximation, we use the Lieb-Robinson bound and set
\be
i {\cal D}^l_s=\int_{|t|\leq l/v_{LR}} {\rm d}t F(t) \Bigl(\partial_s H_s\Bigr)^l(t),
\ee
where
$\Bigl(\partial_s H_s\Bigr)^l(t)$ is supported on the set of sites
within distance $l$ of $X$ and is the approximation to $\exp(i H_s t) \Bigl( \partial_s H_s\Bigr) \exp(-i H_s t)$
given by the Lieb-Robinson bound.
We now bound the difference $\Vert {\cal D}_s-{\cal D}^l_s \Vert$.  This difference is bounded by
\begin{widetext}
\be
\int_{|t|\leq l/v_{LR}} {\rm d}t |F(t)| \Vert 
\Bigl(\partial_s H_s\Bigr)^l(t)-\exp(i H_s t) \Bigl( \partial_s H_s\Bigr) \exp(-i H_s t)\Vert+
\int_{|t|\geq l/v_{LR}} {\rm d}t |F(t)| \Vert \partial_s H_s \Vert.
\ee
Using the Lieb-Robinson bound and Eq.~(\ref{Fdecay}) we arrive at
\be
\label{eq2}
\Vert {\cal D}_s - {\cal D}^l_s \Vert \leq {\rm const.} \Vert \partial_s H_s \Vert (v_{LR}\alpha^2/l) \Bigl(\exp(-l^2/2 \alpha^2 v_{LR}^2)+
g(l) |X|\Bigr).
\ee

Combining Eqs.~(\ref{eq1},\ref{eq2}), and choosing $\alpha=\sqrt{l/v_{LR} \Delta E}$ we find that
\be
\label{Dbound}
\Vert \partial_s P(s)-[i {\cal D}^l_s,P(s)] \Vert \leq {\rm const.} \times (\Vert \partial_s H_s \Vert/\Delta E) \Bigl(\exp(-l \Delta E/2v_{LR})+g(l)\Bigr).
\ee
\end{widetext}

Finally, we define $U(s)$ by $U(0)=I$, the identity operator, and $\partial_s U(s)=i {\cal D}_s U(s)$.  Since $i {\cal D}_s$ is Hermitian by
construction, $U(s)$ is unitary.  Eq.~(\ref{Ubound}) follows from Eq.~(\ref{Dbound}).
\qed

%Unused bibitems

%\bibitem{TV}
%V.\ Turaev and O.\ Viro, 
%Topology \textbf{31}, 865 (1992).
%Unused bibitems

%\bibitem{eno05}
%P.~Etingof, D.~Nikshych, and V.~Ostrik,
%Ann. Math. {\bf 162}, 581 (2005). 
%\bibitem{kr88}
%A.N.~Kirillov and N.Y.~Reshetikhin,
%{\it Representations of the algebra $U_q(sl(2))$, $q$-orthogonal polynomials and invariants of links},
%in V.G.~Kac, ed., {\it Infinite dimensional Lie algebras and groups, Proceedings of the conference held at CIRM, Luminy, Marseille}, p. 285, World Scientific, Singapore (1988).  

\begin{thebibliography}{2}
\bibitem{nonunitarySK} In fact, it is possible to find a braid of length $l$ such that the corresponding matrix $M$ in the representation $\rho \otimes \rho^{*}$ has  a ratio between largest and smallest singular values which grows exponentially with $l$.  This exponential growth actually requires some care to show since the representation is non-unitary, and it does not appear to follow directly from the fact that $\rho$ is dense; we have been able to prove it using a form of the Solovay-Kitaev algorithm.  However, since the particular exponential growth does not play any role in the proof  we omit the details.
\end{thebibliography}
\end{document}